\documentclass[twocolumn,amsmath,amssymb,aps,pre,10pt]{revtex4-1}
\usepackage{graphicx}
\usepackage{nicematrix}
\usepackage{subcaption}
\usepackage{dcolumn}
\usepackage{bm}
\usepackage{float}
\usepackage{multirow}
\usepackage[table,xcdraw]{colortbl}
\usepackage{physics}
\usepackage[hidelinks]{hyperref}
\definecolor{Referee1Comments}{HTML}{FF8000}
\definecolor{Referee2Comments}{HTML}{2A6099}

\usepackage{eurosym}
\usepackage{soul}
\usepackage{dirtytalk}
\usepackage[normalem]{ulem}
\usepackage{amsmath}
\newcommand{\stkout}[1]{\ifmmode\text{\sout{\ensuremath{#1}}}\else\sout{#1}\fi}
 
\begin{document}
\title{A lattice Boltzmann approach for acoustic manipulation}

\author{E. Castro-\'Avila} \email{escastroav@unal.edu.co} 
\affiliation{Simulation of Physical Systems Group, Department of Physics, Universidad Nacional de Colombia, 
Crr 30 \# 45-03, Bogot\'a D.C., Colombia
 }
 
\author{Paolo Malgaretti} \email{p.malgaretti@fz-juelich.de} 
\affiliation{Helmholtz Institute Erlangen-N\"urnberg for Renewable Energy, Forschungszentrum J\"ulich, Cauerstr.\,1, D-91058 Erlangen, Germany}

\author{Jens Harting}
\email{j.harting@fz-juelich.de}
\affiliation{Helmholtz Institute Erlangen-N\"urnberg for Renewable Energy, Forschungszentrum J\"ulich, Cauerstr.\,1, D-91058 Erlangen, Germany}
\affiliation{Department of Chemical and Biological Engineering and Department of Physics, Friedrich-Alexander-Universit\"at Erlangen-N\"urnberg, Cauerstr.\,1, D-91058 Erlangen, Germany} 

\author{J. D. Mu\~noz}\email{jdmunozc@unal.edu.co}
\affiliation{Simulation of Physical Systems Group, Department of Physics, Universidad Nacional de Colombia, 
Crr 30 \# 45-03, Ed. 404, Of. 348, Bogot\'a D.C., Colombia
 }

\date{\today}
\begin{abstract}
We employ a lattice Boltzmann method to compute the acoustic radiation force produced by standing waves on a compressible object for the density matched case. Instead of simulating the fluid mechanics equations directly, the proposed method uses a lattice Boltzmann model that reproduces the wave equation, together with a kernel interpolation scheme, to compute the first order perturbations of the pressure and velocity fields on the object's surface and, from them, the acoustic radiation force. The procedure reproduces with excellent accuracy the theoretical expressions by Gor'kov and Wei for the sphere as the 3D case and an infinitely long cylinder as the 2D case, respectively, even with a modest number of lattice Boltzmann cells. The proposed method shows to be a promising tool for simulating phenomena where the acoustic radiation force plays a relevant role, like acoustic tweezers or the acoustic manipulation of microswimmers, with applications in medicine and engineering.
\end{abstract}


\maketitle

\section{Introduction}\label{sec:intro}
The manipulation of particles by acoustic oscillating fields in aqueous media has won relevance in recent years because of its applications in biotechnology, health, and marine research, among others~\cite{Mohanty2020,Harting2021,Shi2009,Marzo2015}. 
Just like optical tweezers~\cite{Ashkin1970,GKKRWH08}, standing acoustic waves can push objects immersed in a fluid to the nodes (or anti-nodes) of the oscillating field, with many practical advantages. 
Indeed, particles can be manipulated by using acoustic standing waves that only require $\simeq 10^5$ times less power than their optical counterpart~\cite{Ozcelik2018}. Indeed, acoustic tweezers can move objects from 100 nm to 10 mm in size by using acoustic wave intensities between $10^{-2}$ and 10 W/cm$^2$~\cite{n8_Marston1985}. Recently, acoustic transducers were used to extract kidney stones from a pig in non-invasive surgery, demonstrating the potential of acoustical tweezers in medicine~\cite{Ghanem2020}. Many experiments showed the possibility of particle manipulation and confinement, such as those by Lawrence A. Crum~\cite{n3_Crum1971} and Robert E. Apfel~\cite{n3_Apfel1971}, where an agreement with theoretical results was possible~\cite{n3_Marston2023}. In another application,  Zhang et. al. showed how nodal planes of acoustic standing waves can be used as virtual walls to induce a rolling motion of self-assembled microswimmers~\cite{paper_rods}. Acoustic radiation pressure can also be used to study resonant modes on cylindrical capillary bridges, as shown in~\cite{n4_Morse1996} and~\cite{n5_Marr-lyon1997}. More applications for acoustic tweezers can be found in a recent review~\cite{Andreas2017}.

Historically, after the seminal work of Louis V. King~\cite{King1934, King1935}, K. Yosioka and Y. Kawasima in 1955~\cite{n1_Yosioka1955} generalized the analysis for compressible small objects, and L.P. Gor'kov  developed in 1962~\cite{Gorkov1962} a multipole expansion for the calculation of the time-averaged force by a plane standing acoustic wave on a spherical particle immersed in an ideal (non-viscous, isentropic and irrotational) fluid. The non-viscous approximation is valid if the object is large enough (Appendix \ref{ap:viscosity}), and has been successfully used for 60 years \cite{n1_Yosioka1955}.Interestingly, when the wavelength is much larger than the particle size, the acoustic force can be cast into an effective potential, the so-called Gor'kov potential.
The two-dimensional case was studied by Junru Wu, Gonghuan Du et al. in 1990~\cite{WuDu1990} and later by Wei Wei et.al. in 2004~\cite{Wei2004}, who used scattering theory to solve the wave equation for a compressible cylinder of infinite height in a standing wave (For theoretical reviews, see~\cite{n2_Marston2006,Bruus_book,Bruus2011_01,Bruus2012_02,Bruus2012_07}).



All these theoretical studies rely on the assumption that the relevant macroscopic fields, like pressure and velocity, can be written as a perturbative expansion, where the first order contributions $p_1$, for the pressure, and $\vec u_1$, for the velocity, satisfy the wave equation. Although the acoustic radiation force depends on second order contributions, they can be written in terms of $p_1$ and $\vec u_1$, opening the possibility of solving the wave equation \-- instead of the Navier-Stokes equations (NSE) \-- for $p_1$ and $\vec u_1$, and use those results to compute the acoustic radiation force. Because the wave equation can be easily solved with large precision, such a procedure will require less computational effort than a direct NSE simulation (where $p_1$ and $\vec u_1$ are just first order perturbations of the computed pressure and velocity fields), allowing to simulate complex shapes in 2D and 3D as required for today's microfluidics applications.

Here, we show how to use a 
lattice Boltzmann method (LBM) that simulates the wave equation~\cite{ChopardArticle} to compute the acoustic radiation force on an object immersed in an inviscid fluid. 
Our approach reproduces with great accuracy the theoretical predictions by Gor'kov~\cite{Gorkov1962} and Wei~\cite{Wei2004} at much lower computational costs than previous LBM approaches simulating the full Navier-Stokes equations~\cite{cosgrove2004,Haydock2005}.
The remainder of this paper is organized as follows: Sec.~\ref{sec:general_arf} presents the acoustic radiation force as derived by  King~\cite{King1934}. Next, Sec.~\ref{sec:ARF_Gorkov_3D} reviews the theoretical deduction of Gor'kov's potential, together with an analogous development by us for the 2D case in Sec.
`\ref{sec:ARF_Gorkov_2D}. Then, Chopard's LBM model for waves is briefly explained in Sec.~\ref{sec:LBW}, with some benchmarks to validate the implementation in our simulation code. Our results for the acoustic radiation force in the 3D case, where the particle is defined from now on as a sphere and the acoustic radiation force-per-length in the 2D case for an infinitely long cylinder, are reported in Sec.~\ref{sec:results}, showing that the computed forces follow the theoretical predictions with excellent accuracy.

\section{The second order acoustic radiation force}\label{sec:general_arf}
The results in this section were first developed by L. V. King \cite{King1934}. Hereby, we present a summary of that work. Consider an inviscid and compressible fluid in an irrotational flow with pressure $P$, density $\rho$, velocity $\vec u$, and momentum $\vec J=\rho\vec u$. The fluid is described by the mass conservation law,
\begin{equation}\label{eq:mass_cons}
    \frac{\partial\rho}{\partial t} + \vec\nabla\cdot\vec J = 0\quad,
\end{equation}
and Euler's equation, which can be rewritten as a momentum conservation law,
\begin{equation}\label{eq:euler_vec}
    \frac{\partial\vec J}{\partial t}=-\vec\nabla P+\vec\nabla\cdot\left(\vec u \otimes \vec J \right) = -\nabla \cdot \Pi\quad.
\end{equation}
Here, $\Pi$ is the momentum flux density tensor, described as
\begin{equation}\label{eq:flux}
    \Pi=\mathbb{I}P+\vec u \otimes \vec J \quad.
\end{equation}
By second Newton's law, the force $\vec F$ acting on a static object immersed in the fluid equals the momentum exchange of the fluid across a volume control around the object~\cite[app.~A]{Manneberg2009}
\begin{equation}\label{eq:ARF_general}
    \vec F = \int \frac{\partial\vec J}{\partial t} dV = - \oint \Pi\cdot\vec{dS}
\end{equation}

Acoustic waves propagate in the fluid as small variations in pressure, density and velocity around steady values. So, the three fields can be written as follows:
\begin{equation}\label{eq:expand_1}
    \rho = \rho_0 + \rho_1\quad,\quad P = p_0 + p_1\quad,\quad\vec u = \vec 0 + \vec u_1\quad,
\end{equation}
where $p_1$, $\rho_1$ and $\vec u_1$ are first order perturbations mentioned in Sec.~\ref{sec:intro}, and we assume  $u_0=0$ (that is, the fluid is at rest). Since acoustic waves in a fluid are adiabatic compressions \cite{Elmore}, the first order pressure and density fluctuations are related as \cite[p.~120]{Kinsler}
\begin{equation}\label{eq:eq_state}
    p_1 = c_0^2\rho_1\quad,
\end{equation}
where $c_0^2$ is the adiabatic speed of sound in the fluid. By taking  first order contributions only (as in Eq.~\eqref{eq:expand_1}), Eqs.~\eqref{eq:euler_vec} and \eqref{eq:mass_cons} are linearized as
\begin{subequations}\label{eq:linear_NSE_eqs}
\begin{equation}\label{eq:linear_momentum_conserv}
    \rho_0\frac{\partial\vec u_1}{\partial t} +\nabla p_1 = 0\quad,
\end{equation}
\begin{equation}\label{eq:linear_mass_conserv}
    \frac{\partial\rho_1}{\partial t} + \rho_0\nabla\cdot\vec u_1 = 0\quad.
\end{equation}    
\end{subequations}
By combining Eqs.~\eqref{eq:linear_NSE_eqs}, 
and \eqref{eq:eq_state}, one obtains that $p_1$ fulfills a wave equation,
\begin{equation}\label{eq:waves_pressure}
    \frac{1}{c_0^2}\frac{\partial^2 p_1}{\partial t^2} = \nabla^2 p_1\quad.
\end{equation}
Since the flow is irrotational, the velocity $\vec u=\vec u_1$ can be written as the gradient of a scalar velocity potential $\phi$, and Eq.~\eqref{eq:linear_momentum_conserv} transforms to give
\begin{equation}\label{eq:grad_phi_u1_p1}
    \vec u_1 = \nabla\phi\quad\mbox{,}\quad p_1 = -\rho_0\frac{\partial\phi}{\partial t}\quad.
\end{equation}
By replacing Eq.~\eqref{eq:grad_phi_u1_p1} into Eq. \eqref{eq:linear_mass_conserv}, and using Eq.~\eqref{eq:eq_state}, we obtain that the velocity potential also fulfills a wave equation, 
\begin{equation}\label{eq:waves_phi}
    \frac{1}{c_0^2}\frac{\partial^2\phi}{\partial t^2} = \nabla^2 \phi\quad.
\end{equation}
In most experiments, the time scale of the object's motion is much larger than the oscillation period, and, thus, only the time average of the acoustic radiation force $\langle\vec F\rangle$ is relevant for particle's dynamics. If only first order fluctuations were considered \eqref{eq:expand_1}, acoustic radiation force \eqref{eq:ARF_general} would reduce to
\begin{equation}\label{eq:ARF_1_order}
    \vec F = - \oint P \vec{dS}\quad,
\end{equation}
because the term $\vec u \otimes \vec J$ is second order.
Since the pressure perturbation oscillates harmonically in time, $p_1\propto e^{i\omega t}$, the time average of \eqref{eq:ARF_1_order} automatically vanishes, and there is no first order contribution to the time-averaged force. 
The origin of that force lies, therefore, in higher-order contributions.
By extending the fluctuations up to second order,
\begin{subequations}\label{eq:expand_2}
\begin{equation}\label{eq:expand_rho_2}
    \rho = \rho_0 + \rho_1 + \rho_2\quad,
\end{equation}
\begin{equation}\label{eq:expand_p_2}
    P = p_0 + p_1 + p_2\quad\text{ and}
\end{equation}    
\begin{equation}\label{eq:expand_u_2}
    \vec u = \vec 0 + \vec u_1 + \vec u_2\quad,
\end{equation}
\end{subequations}
one obtains that the gradient for the perturbations in the pressure field can be written in terms of first order contributions only (please see App.~\ref{ap:arf_second} for details) as
\begin{equation}\label{eq:euler_king_grad}
    \nabla(P - p_0) = \nabla\left(\frac{p_1^2}{2\rho_0 c_0^2} - \frac{\rho_0}{2}u_1^2 - \rho_0\frac{\partial\phi}{\partial t}\right)\quad ,
\end{equation}
leading to the following expression for the acoustic radiation force, written in index notation:
\begin{equation}\label{eq:arf_integral}
    \langle F_{i} \rangle = - \oint \left\langle\left(-\rho_0\frac{u_1^2}{2}+\frac{p_1^2}{2\rho_0c_0^2}\right)\delta_{ij}+\rho_0 u_{1i}u_{1j}\right\rangle dS_j \quad,
\end{equation}
where the triangular brackets $\left< \cdot\right>$ denote time averages.
The contribution of the term $\rho_0\frac{\partial\phi}{\partial t}$ in Eq.~\eqref{eq:euler_king_grad} vanishes because the velocity potential $\phi\propto e^{i\omega t}$ is also harmonic. 
Eq.~\eqref{eq:arf_integral} allows us to compute the time-averaged acoustic radiation force (or force-per-length in the 2D case) on an immersed object from the pressure ($p_1$) and velocity ($\vec u_1$) fields obtained by simulating the acoustic waves directly (Eqs.~\eqref{eq:waves_pressure} and \eqref{eq:linear_momentum_conserv}), instead of solving the full Navier-Stokes equations.



\section{Acoustic radiation force on a sphere}\label{sec:ARF_Gorkov_3D}

\begin{figure}
    \centering
    \includegraphics[width=0.45\textwidth]{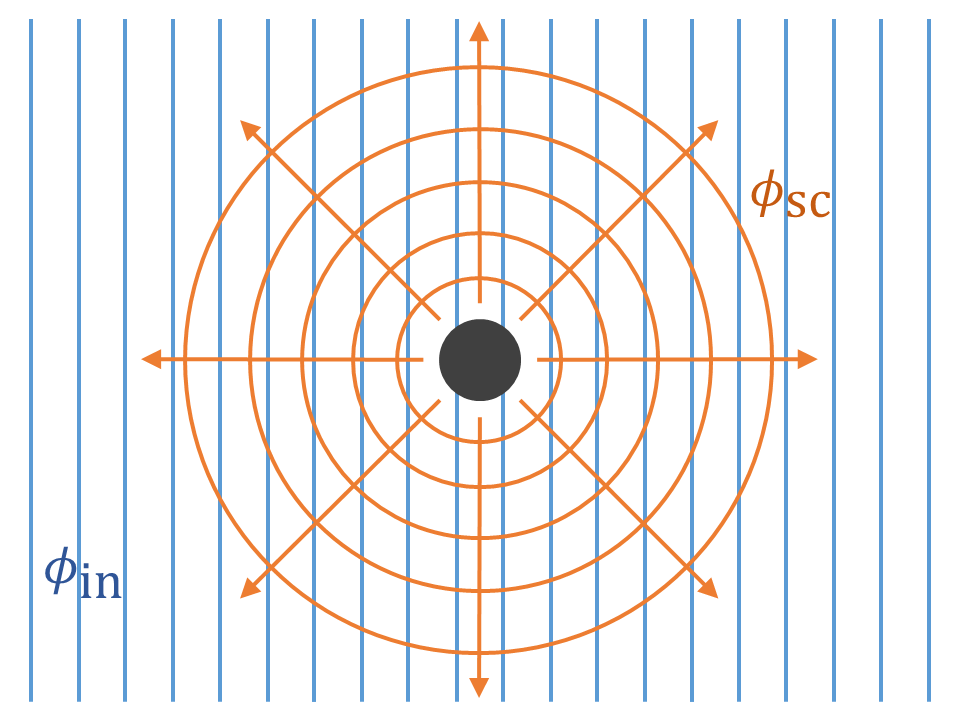}
    \caption{Decomposition into incident and scattering fields.}
    \label{fig:scattering}
\end{figure}

The results presented in this section were deducted by K. Yosioka and Y. Kawasima ~\cite{n1_Yosioka1955} by using a Hankel functions expansion, and by L. P. Gor'kov via a multipole expansion ~\cite{Gorkov1962,Manneberg2009}. We aim to compute the average force by acoustic waves acting on a spherical object immersed in a liquid. Let us assume that the object's radius $R_p$ is much smaller than the wavelength $\lambda$, i.e. $R_p\ll\lambda$ or $kR_p\ll 1$. Also, it is assumed that the viscous penetration depth $\delta=\sqrt{2\nu/\omega}$, with $\nu$ the kinetic viscosity, is much smaller than $R_p$, so that the condition $\delta/R_p\ll1$ is satisfied. Under such condition, viscosity can be neglected (For a more detailed discussion on the effects of viscosity, see Appendix \ref{ap:viscosity} ). Note that larger values of $kR_p$ would also break the validity of the presented model. Improved models on the acoustic radiation force using  expansions to greater orders in $kR_p$ are found in~\cite{n6_Marston2017} and~\cite{n7_Marston2017}.
With this approximation, it is possible to solve Eq.~\eqref{eq:waves_phi} by dividing the potential, pressure and velocity fields into an incident and a scattered part as shown in Fig.~\ref{fig:scattering}~\cite{Gorkov1962}, that is
\begin{equation}\label{eq:in_sc_potential}
    \phi = \phi_{\text{in}} + \phi_{\text{sc}}
    \quad,\quad    
    p_1 = p_{\text{in}} + p_{\text{sc}}
    \quad,\quad
    \vec u_1 = \vec u_{\text{in}} + \vec u_{\text{sc}} \mbox{.}
\end{equation}
The incident fields $(in)$
are the solution for the ongoing waves as if there was no spherical particle, while the scattered fields $(sc)$
are the differences between the total fields and the incident ones. By replacing \eqref{eq:in_sc_potential} into \eqref{eq:arf_integral}, three main terms appear: The first one only depends on the incident fields and it does not contribute to the force; the second one depends on the scattered field $\phi_{\text{sc}}$ only, which will be proportional to $R_p^6$ and, thus, vanishes because of $R_p\ll\lambda$. 
The last and only surviving term depends on both $\phi_{\text{in}}$ and $\phi_{\text{sc}}$, and the force reduces to (see App.~\ref{ap:in_sc_force} for details)
\begin{equation}\label{eq:F_sc}
    \langle \vec F \rangle = - \rho_0\int \left\langle \vec u_{\text{in}}\left(\nabla^2\phi_{\text{sc}}-\frac{1}{c_0^2}\frac{\partial^2\phi_{\text{sc}}}{\partial  t^2}\right)\right\rangle dV\quad.
\end{equation}
The term between parenthesis is not zero because the object acts as a local source for the scattered field. Thus, a retarded-time multipolar expansion is proposed as a solution for $\phi_{\text{sc}}$,
\begin{align}\label{eq:scattered}
    \phi_{sc} &= -\frac{a(t')}{r'}-\frac{\vec A(t') \cdot \vec r'}{r'^3} + \dots\nonumber\\
    &\approx \phi_{\text{mp}} + \phi_{\text{dip}}
\end{align}
where $\vec r'=\vec r-\vec v t$ is the position of the center of the sphere, moving with velocity $\vec v$, and $t'=t-r/c$ is a retarded time (with $t'\approx t$ close to the object)~\cite{Landau,Jackson}. If $kR_p$ is greater enough, more terms should be considered in \eqref{eq:scattered} in this expansion. Corrections for larger $kR_p$ are not reviewed in this work, but the reader should refer to~\cite{n6_Marston2017,n7_Marston2017}.
\begin{figure}[b]
    \centering
    \includegraphics[width=0.45\textwidth]{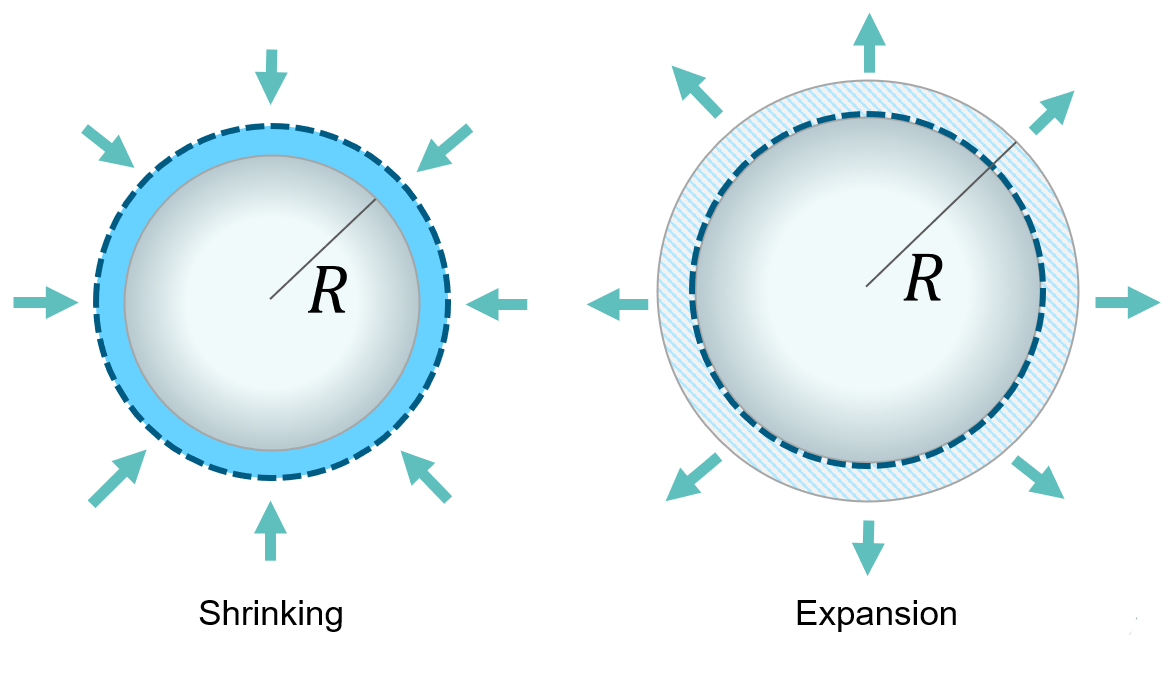}
    \caption{The object's pulsating expansions and contractions induce spherical waves coinciding with the monopole term of the scattered velocity potential field.}
    \label{fig:mp_term}
\end{figure}

The monopole term $\phi_{mp}=a(t)/r`$ can be found by assuming that the particle is able to compress and expand isotropically as shown in figure \ref{fig:mp_term}, changing its volume $V_p$ in response to the incident perturbations $p_{\text{in}}$ as
\begin{equation}\label{eq:pressure_bulk_3D}
    dp_{in} = -B_p \frac{dV_p}{V_p}\quad\mbox{, or}\quad
    \frac{\partial p_{in}}{\partial t} = -\frac{B_p}{V_p} \frac{dV_p}{dt}\quad,
\end{equation}
with $B_p=c_p^2\rho_p$ the particle's bulk modulus, $c_p$ the particle's speed of sound and $\rho_p$ the particle's density. Thus, by using \eqref{eq:eq_state},
\begin{equation}\label{eq:VpTimeDerivative}
    \frac{d V_p}{d t} =  -V_p\frac{c_0^2}{\rho_pc_p^2}\frac{\partial\rho_{\text{in}}}{\partial t}\quad .
\end{equation}
Now, let us consider a spherical region $\Omega$ of radius $R_{\Omega}$ concentric to the sphere of radius $R_p$, with  $\lambda\gg R_{\Omega} > R_p$. When the sphere expands, the mass flux leaving $\Omega$ through its surface $\partial\Omega$ equals the rate at which the sphere pushes fluid out of it,
\begin{align}\label{eq:mass_flux_3D}
    \dot{m} &= \frac{\partial}{\partial t}[(\rho_0 + \rho_{\text{in}})V_p]\\
    &= \int_{\Omega}\dot{\rho}dV = \oint_{\partial\Omega} (\rho_0\grad\phi_{\text{mp}})\cdot\hat r dS = 4\pi a(t)\rho_0\quad ,
\end{align}
with $\hat r$ the radial unitary vector. By
\eqref{eq:pressure_bulk_3D} and \eqref{eq:VpTimeDerivative} the total derivative in \eqref{eq:mass_flux_3D} becomes
\begin{equation}
    \dot{m} = -V_p\frac{c_0^2}{\rho_pc_p^2}\frac{\partial}{\partial t}\left(\rho_0\rho_{\text{in}} + \frac{1}{2}\rho_{\text{in}}^2\right) + V_p\frac{\partial\rho_{\text{in}}}{\partial t}\mbox{ .}
\end{equation}
Because the incident density is a perturbation of the total density of the fluid (i.e. $\rho_0\gg\rho_{\text{in}}$) the term $\frac{1}{2}\rho_{\text{in}}^2$ is much smaller than $\rho_0\rho_{in}$ and can be vanished. Therefore,
\begin{equation}\label{ap.arf.eq:m_in_solved}
    \dot{m} =4\pi a(t)\rho_0= V_p\frac{\partial\rho_{\text{in}}}{\partial t}\left(1-\frac{\rho_0c_0^2}{\rho_pc_p^2}\right)\quad,
\end{equation}
and $a(t)$ turns out to be 
\begin{equation}\label{eq:a_3D}
    a(t) = \frac{R_p^3}{3\rho_0}f_1\frac{\partial\rho_{\text{in}}}{\partial t}
    \quad\mbox{, with}\quad
    f_1 = 1-\frac{\rho_0 c_0^2}{\rho_pc_p^2}\quad.
\end{equation}

\begin{figure}[b]
    \centering
    \includegraphics[width=0.45\textwidth]{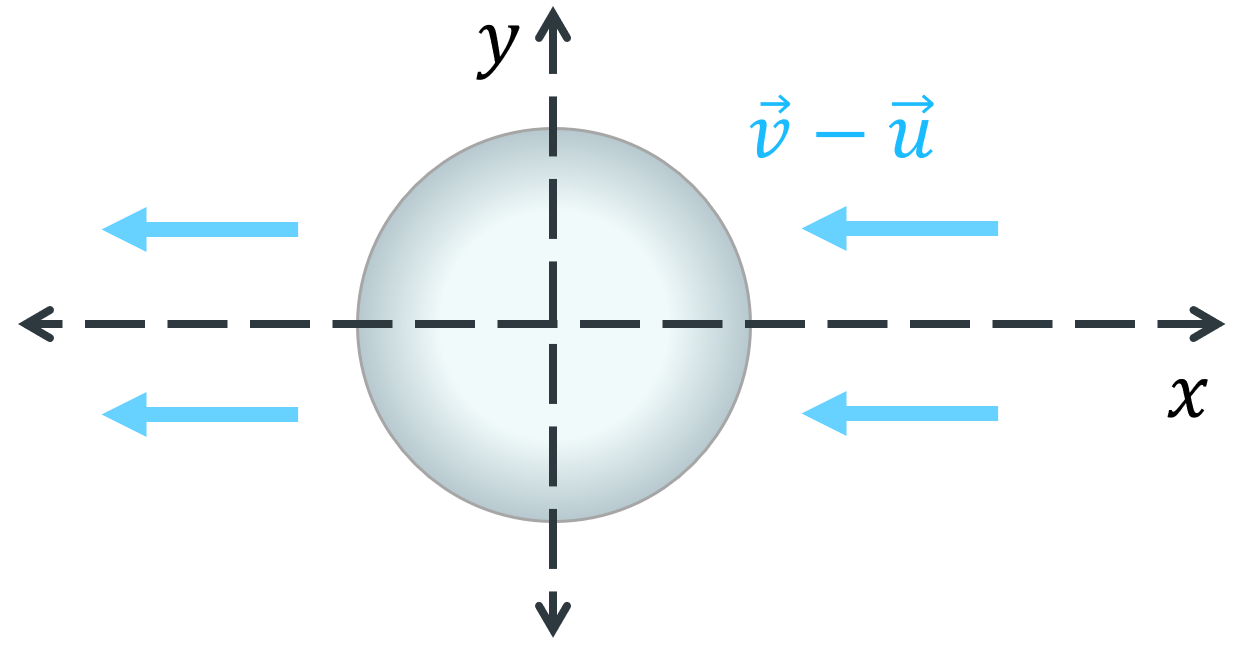}
    \caption{The accelerated motion of the object pushes the surrounding fluid and induces an \textit{added mass}, i.e. a drag force proportional to the acceleration. That force allows to compute the dipole term of the scattered velocity potential field.}
    \label{fig:dip_term}
\end{figure}

The dipole term 
\begin{equation}\label{eq:DipolePotential}
    \phi_{\text{dip}}=-\vec A\cdot\vec r'/r'^3
\end{equation}
comes into play when the sphere moves with a changing velocity $\vec v$ in a fluid with velocity $\vec u_1$. The velocity potential for this situation is the sum of the one for a sphere at rest in a fluid approaching with velocity $\vec u_1$ plus the one of a sphere moving with velocity $\vec v$ in a fluid at rest~\cite{Kundu},
\begin{equation}\label{eq:VelocityPotentialForAMovingSphere_1}
        \phi=\left(1+\frac{R_p^3}{2r`^3}\right)\left(\vec u_1\cdot\vec r'\right)
            -\frac{R_p^3}{2r`^3}\left(\vec v\cdot\vec r'\right)\quad.
\end{equation}
The potential can be rearranged as
\begin{equation}\label{eq:VelocityPotentialForAMovingSphere_2}
        \phi=-\frac{R_p^3}{2r`^3}\left(\vec v-\vec u_1\right)\cdot\vec r'
            +\left(\vec u_1\cdot\vec r'\right)\quad,
\end{equation}
where the last term is easily identified as the incident potential $\phi_{in}$, and the first term, as the one of a sphere moving in a fluid at rest with velocity $\left(\vec v-\vec u_1\right)$ (see Fig. \ref{fig:dip_term}). Indeed, by comparing $\phi_{in}$ in \eqref{eq:VelocityPotentialForAMovingSphere_2} with \eqref{eq:DipolePotential}, one can identify
\begin{equation}\label{eq:VectorPotentialA}
        \vec A(r')=-\frac{R_p^3}{2}\left(\vec v-\vec u_1\right)\quad.
\end{equation}

On the one side and according to~\cite{Kundu}, the force on the sphere produced by that potential is
\begin{equation}\label{eq:AparentMassForce}
        F_{sc}=-M_a\left(\frac{\partial \vec v}{\partial t}-\frac{\partial \vec u_1}{\partial t}\right)\quad,
\end{equation}        
   with
\begin{equation}
        M_a=\frac{2}{3}\pi R_p^3\rho_0
\end{equation}
the \textit{added mass} for the sphere.
On the other hand, if the sphere was made of fluid, it would move with velocity $\vec v=\vec u_1$ and, therefore, the force on the sphere from a potential $\phi=\vec u_1\cdot\vec r'$ must be exactly the one we need to move that fluid sphere with an acceleration $\frac{\partial\vec u_1}{\partial t}$
\begin{equation}\label{eq:IncidentForce}
    F_{in}=M_f\frac{\partial\vec u_1}{\partial t}\quad\mbox{, with}\quad M_f=\frac{4}{3}\pi R_p^3\rho_0 \quad.
\end{equation}
The total force acting on the sphere of density $\rho_p$ and mass $M_p=4\pi R_p^3/3\rho_p$ is the sum of those two forces. Thus,
\begin{equation}\label{eq:TotalForce}
    M_p\frac{\partial \vec v}{\partial t}=
    -M_a \left(\frac{\partial \vec v}{\partial t}-\frac{\partial \vec u_1}{\partial t}\right)
    +M_f\frac{\partial\vec u_1}{\partial t}\quad.
\end{equation}
By replacing the expressions for the masses, we obtain
\begin{equation}\label{eq:vwsu1_1}
    \left(\rho_p+\frac{\rho_0}{2}\right)\frac{\partial \vec v}{\partial t}=
    \frac{3}{2}\rho_0\frac{\partial\vec u_1}{\partial t}
    \quad.
\end{equation}
If we assume that $\vec v=0$ when $\vec u_1=0$, integrating with respect to time $t$ gives us
\begin{equation}\label{eq:v}
    \vec v=\frac{3\rho_0}{2\rho_p+\rho_0}\vec u_1\quad.
    \quad
\end{equation}
and the vector field $\vec A(t)$ becomes
\begin{equation}\label{eq:A_3D}
    A(t) = \frac{R_p^3}{2}f_2\vec u_1
    \quad\mbox{, with}\quad
    f_2 = \frac{2(\rho_p-\rho_0)}{2\rho_p+\rho_0}\quad.
\end{equation}

By plugging \eqref{eq:a_3D} and \eqref{eq:A_3D} into 
\eqref{eq:scattered} and  \eqref{eq:F_sc}, and by solving the integration (see Appendix \ref{ap:integrate_gorkov} for details), we have
\begin{equation}\label{eq:F_grad_U}
    \langle F_i \rangle = -\partial_i V_p\left(f_1\frac{1}{2\rho_0 c_0^2}\langle p'^2_{\text{in}} \rangle + \frac{3}{4}\rho_0 f_2\langle v^2_{\text{in}} \rangle\right) = -\nabla U
\end{equation}
where the potential $U$ is defined as
\begin{equation}
    U = V_p\left(f_1\frac{1}{2\rho_0 c_0^2}\langle p'^2_{\text{in}} \rangle + \frac{3}{4}\rho_0 f_2\langle v^2_{\text{in}} \rangle\right)\quad .
\end{equation}
That is the Gor'kov potential. This potential is commonly used for acoustic levitation of small objects, even regardless of the object`s shape as soon as $R_0 \ll \lambda$ is satisfied. In the particular case of incident stationary waves,
\begin{equation}\label{eq:p_in_standing}
    p_1(x,t) = P_a\sin(\omega t)(\cos kx)\quad,
\end{equation}
with $P_a$ the pressure amplitude, Eq. \eqref{eq:grad_phi_u1_p1} give us the velocity as
\begin{equation}\label{eq:u_in_standing}
    u_{1x}(x,t) = -\frac{P_a}{c_0\rho_0}\cos\omega t\sin kx\quad.
\end{equation}
By plugging \eqref{eq:p_in_standing} and \eqref{eq:grad_phi_u1_p1} into the Gor'kov potential and solving the time-average integration, this potential for standing waves becomes
\begin{equation}\label{eq:Gorkov_potential_3D}
    U = \frac{V_pP_a^2}{4\rho_0 c_0^2}\left(f_1\cos^2kx+\frac{3}{2}f_2\sin^2kx\right)\quad.
\end{equation}
The force can be gathered using \eqref{eq:F_grad_U}, \eqref{eq:a_3D} and \eqref{eq:A_3D}, leading to
\begin{equation}\label{eq:ARF_Gorkov_3D}
    F_x = -\frac{\pi R_p^3P_a^2k}{3\rho_0 c_0^2}\Phi(\tilde\rho,\tilde\kappa)\sin 2kx\quad,
\end{equation}
where $\tilde\rho = \rho_p/\rho_0$ , $\tilde\kappa = \kappa_p/\kappa_0$ and $\Phi(\tilde\rho,\tilde\kappa)$ is defined as
\begin{equation}
    \Phi(\tilde\rho,\tilde\kappa) = \frac{5\tilde\rho -2}{2\tilde\rho+1}-\tilde\kappa\quad.
\end{equation}
This is the theoretical result to compare with in 3D. 

\section{Acoustic radiation force per length on a cylinder}\label{sec:ARF_Gorkov_2D}

The results of this section were developed by Wei \cite{Wei2004}. Nevertheless, instead of following the deduction by Wei, let us develop the same procedure as in the previous section, but in two dimensions, a procedure that will give the same results. As in \eqref{eq:F_sc}, the acoustic radiation force-per-length is computed from the scattered potential as 
\begin{equation}\label{eq:F_sc_2D}
    \langle F_{i} \rangle = - \rho_0\int \left\langle u_{\text{in}}^i\left(\nabla^2\phi_{sc}-\frac{1}{c_0^2}\frac{\partial^2\phi_{sc}}{\partial  t^2}\right)\right\rangle dA\quad{.}
\end{equation}
Because the cylinder is infinite in height, \eqref{eq:F_sc_2D} computes a force per unit length since the cylinder would get an infinite surface and the integration is done along the circular boundary of its cross-section. A retarded time multipolar expansion is also a good approximation in 2D,
\begin{equation}\label{eq:scattered_2D}
    \phi_{sc} = -a(t')\log{r} - \frac{\vec A(t')\cdot\vec r'}{r'^2} + \dots
\end{equation}
with $\vec r'=\vec r-\vec v t$ and $t'=t-r/c\approx t$. 

As in the 3D case, the scalar field $a(t)$ in the monopole term $\phi_{mp}=-a(t')\log{r}$ can be found by assuming an isotropic compression and expansion of the cylinder due to the incident field $p_{\text{in}}$ by 
\begin{equation}\label{eq:pressure_bulk_2D}
    dp_{in} = -B_p^{\text{2D}} \frac{dA_p}{A_p}\quad\mbox{, or}\quad
    \frac{\partial p_{in}}{\partial t} = -\frac{B_p^{\text{2D}}}{A_p} \frac{dA_p}{dt}\quad,
\end{equation}
with $A_p$ the cylinder's area, $B_p^{\text{2D}}=-AdP/dA=c_p^2\rho_p$ the particle's bulk modulus in 2D. Furthermore, since $p_{in}=\rho_{in} c_0^2$, with $c_0$ the speed of sound in the fluid,
\begin{equation}\label{eq:ApTimeDerivative_ap}
    \frac{d A_p}{d t} =  -A_p\frac{c_0^2}{\rho_pc_p^2}\frac{\partial\rho_{\text{in}}}{\partial t}\quad.
\end{equation}
Consider a circular region $\Sigma$ of radius $R_{\Sigma}$ concentric to the cylinder with  $\lambda\gg R_{\Sigma} > R_p$. On the one side, the mass flux leaving $\Sigma$ through its boundary $\partial\Sigma$ equals the rate at which the cylinder pushes fluid out of it 
\begin{equation}\label{eq:MassBalance_ap_2D}
    \dot{m} = \oint_{\partial\Sigma} (\rho_0\vec\nabla\phi_{mp})\cdot\hat r dl
    = 2\pi a(t)\rho_0 \quad,
\end{equation}
On the other hand, 
\begin{equation}\label{eq:MassBalance_totaldV_2D}
    \dot{m}  = \frac{\partial}{\partial t}[(\rho_0 + \rho_{\text{in}})A_p] \approx A_p\frac{\partial\rho_{\text{in}}}{\partial t}\left(1-\frac{\rho_0c_0^2}{\rho_pc_p^2}\right)\mbox{ ,}
\end{equation}
where we used \eqref{eq:pressure_bulk_2D} and \eqref{eq:ApTimeDerivative_ap}. By equaling \eqref{eq:MassBalance_ap_2D} and \eqref{eq:MassBalance_totaldV_2D} by 
taking into account once again that $\frac{1}{2}\rho_{\text{in}}^2$ is much smaller than $\rho_0\rho_{\text{in}}$, we finally get
\begin{equation}\label{eq:a_2D}
        a(t) = \frac{R_p^2}{2\rho_0}\frac{\partial\rho_{in}}{\partial t}f_1
    \quad\mbox{, with}\quad
    f_1=1-\frac{\kappa_p}{\kappa_0}\quad .
\end{equation}

Like in the 3D case the velocity potential around a cylinder traveling with velocity $\vec v$ in a fluid moving with velocity $\vec u_1$ is \cite{Kundu}
\begin{equation}\label{eq:VelocityPotentialForAMovingDisk_1}
        \phi=\left(1+\frac{R_p^2}{r`^2}\right)\left(\vec u_1\cdot\vec r'\right)
            -\frac{R_p^2}{r`^2}\left(\vec v\cdot\vec r'\right)\quad,
\end{equation}
that can be rearranged as
\begin{equation}\label{eq:VelocityPotentialForAMovingDisk_2}
        \phi=-\frac{R_p^2}{r`^2}\left(\vec v-\vec u_1\right)\cdot\vec r'
            +\left(\vec u_1\cdot\vec r'\right)\quad.
\end{equation}
Here, again, the last term is the incident potential $\phi_{in}$, and the first term, the one of a cylinder moving in a fluid at rest with velocity $\left(\vec v-\vec u_1\right)$. By comparing $\phi_{in}$ in \eqref{eq:VelocityPotentialForAMovingDisk_2} with the dipole potential $\phi_{dip}=- \frac{\vec A(t)\cdot\vec r'}{r'^2}$ we identify
\begin{equation}\label{eq:VectorPotentialA_2D}
        \vec A(r')=-R_p^2\left(\vec v-\vec u_1\right)\quad.
\end{equation}

The force on the cylinder produced by that potential is
\begin{equation}\label{eq:AparentMassForce2D}
        F_{sc}=-M_a\left(\frac{\partial \vec v}{\partial t}-\frac{\partial \vec u_1}{\partial t}\right)\quad\mbox{, with}\quad
        M_a=\pi R_p^2\rho_0
\end{equation}
the \textit{added mass} for the cylinder \cite{Bernard2015}.

As in the 3D case, if the cylinder were made of fluid, it would move with velocity $\vec v=\vec u_1$ and, therefore, the force on the cylinder from a potential $\phi=\vec u_1\cdot\vec r'$ must be exactly the one we need to move that fluid cylinder with an acceleration $\frac{\partial\vec u_1}{\partial t}$
\begin{equation}\label{eq:IncidentForce2D}
    F_{in}=M_f\frac{\partial\vec u_1}{\partial t}\quad\mbox{, with}\quad M_f=\pi R_p^2\rho_0 \quad.
\end{equation}
The total force-per-length acting on the cylinder of density $\rho_p$ and mass $M_p=\pi R_p^2\rho_p$ is the sum of those two contributions. Thus,
\begin{equation}\label{eq:TotalForce2D}
    M_p\frac{\partial \vec v}{\partial t}=
    -M_a \left(\frac{\partial \vec v}{\partial t}-\frac{\partial \vec u_1}{\partial t}\right)
    +M_f\frac{\partial\vec u_1}{\partial t}\quad.
\end{equation}
By replacing the expressions for the masses, we obtain
\begin{equation}\label{eq:vwsu1_1_2D}
    \left(\rho_p+\rho_0\right)\frac{\partial \vec v}{\partial t}=
    2\rho_0\frac{\partial\vec u_1}{\partial t}
    \quad.
\end{equation}
If we assume that $\vec v=0$ when $\vec u_1=0$, integrating with respect to time $t$ gives us
\begin{equation}\label{eq:v_2D}
    \vec v=\frac{2\rho_0}{\rho_0+\rho_p}\vec u_1\quad.
    \quad
\end{equation}
and the vector field $\vec A(t)$ becomes
\begin{equation}\label{eq:A_2D}
    A(t) = R_p^2f_2\vec u_{\text{in}}
    \quad\mbox{, with}\quad
    f_2 = \frac{\rho_0-\rho_p}{\rho_0+\rho_p}\quad.
\end{equation}

After plugging in \eqref{eq:a_2D} and \eqref{eq:A_2D} into \eqref{eq:F_sc_2D} and using the same standing incident wave written in \eqref{eq:p_in_standing} and \eqref{eq:u_in_standing}, the two-dimensional Gor'kov's potential takes the form deducted by Wei et al.
\begin{equation}
    U = \frac{A_pp_0^2}{4\rho_0 c_0^2}\left(f_1\cos^2kx+f_2\sin^2kx\right)
\end{equation}
and the acoustic radiation force-per-length is
\begin{equation}\label{eq:ARF_Gorkov_2D}
    F_x = -\frac{\pi R_p^2P_a^2k}{4\rho_0 c_0^2}\Phi(\tilde\rho,\tilde\kappa)\sin 2kx\quad,
\end{equation}
where $\tilde\rho = \rho_p/\rho_0$, $\tilde\kappa = \kappa_p/\kappa_0$ and $\Phi(\tilde\rho,\tilde\kappa)$ is defined as
\begin{equation}
    \Phi(\tilde\rho,\tilde\kappa) = \frac{3-\tilde\rho}{1+\tilde\rho}-\tilde\kappa\quad.
\end{equation}
This last expression of the force-per-length \eqref{eq:ARF_Gorkov_2D} coincides with the equations 18 and 22 of the paper of Wei et al.~\cite[page.~204]{Wei2004},validating this shorter deduction.

\section{Lattice-Boltzmann for acoustics}\label{sec:LBW}

\begin{figure}
    \begin{subfigure}{0.24\textwidth}
    \centering
    \includegraphics[width=\textwidth]{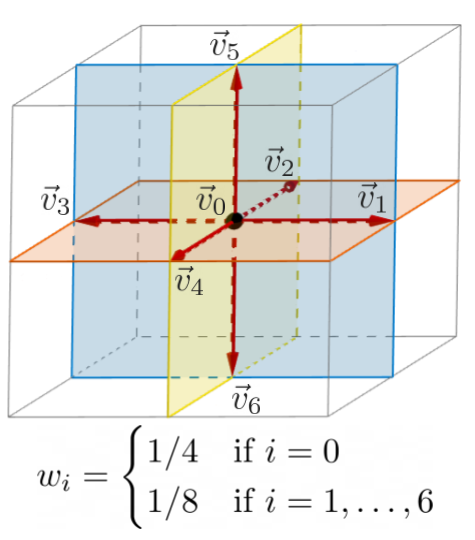}
    \caption{}
    \label{fig:d3q}
    \end{subfigure}
    \begin{subfigure}{0.2\textwidth}
    \centering
    \includegraphics[width=\textwidth]{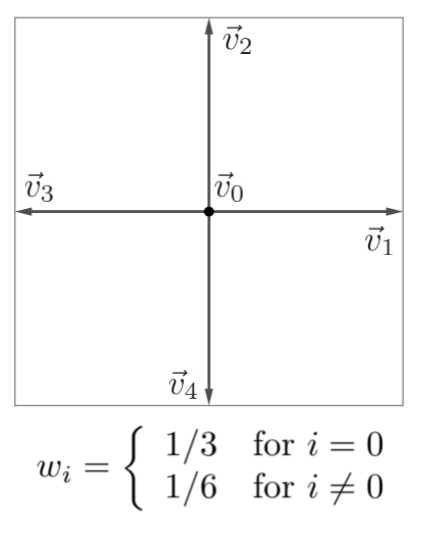}
    \caption{}
    \label{fig:d3q19}
    \end{subfigure}
    \caption{D3Q7 (a) and D2Q5 (b) velocity sets and their corresponding weights.}
    \label{fig:velocity_sets}
\end{figure}

In an LBM, space is divided into cells, and a set of velocity vectors $\vec v_i$ allows for transport from each cell to the neighboring ones (Fig.~\ref{fig:velocity_sets}).
At each cell position $\vec x$ and associated with each velocity vector $\vec v_i$, there is at least one variable $f_i(\vec x,t)$, called a \textit{distribution function}, which represents the density of some hypothetical particles there, moving with that velocity and transporting information from cell to cell. At each time step, the $f_i$ are, first, mixed inside every cell into new values $f_i^{\rm new}$ by using Boltzmann's transport equation (collision) and, then, transported to the neighboring cells along the vector they are associated with (advection). Because all the information for the collision is inside the cell itself and the advection is blind, all cells can evolve each time step in a fully independent way. That makes the LBM suitable to run in parallel on multiple CPUs or graphic cards~\cite{Kruger,HVC04}.

In an LBM for acoustics \cite{ChopardArticle}, the fluid density $\rho_1$, the  momentum $\vec J_1=\rho_0\vec u_1$ and the pressure $p_1$ at any cell are computed as
\begin{equation}\label{eq:MacroscopicFields}
    \rho_1 = \sum_i f_i\quad,\quad
    \vec J_1 = \sum_i \vec v_i f_i\quad,\quad
    p_1 = c^2\rho_1\quad,
\end{equation}
where $c$ is the speed of sound for the fluid.
At each time step $\Delta t$ every cell computes new values $f_i^{\rm new}$ for the distribution functions by using Boltzmann's transport equation in the BGK (Bhatnagar-Gross-Krook) approximation,
\begin{equation}\label{eq:LBM-Collision}
    f_i^{\rm new}(\vec x,t)=\left(1-\frac{\Delta t}{\tau}\right)f_i+\left(\frac{\Delta t}{\tau}\right)f_i^{eq}\quad\mbox{, with}\quad \tau=\frac{1}{2}\mbox{,}
\end{equation}
where the equilibrium values $f_i^{eq}$ are computed from the macroscopic fields \eqref{eq:MacroscopicFields}. In contrast with LBM schemes for fluids, $\tau=1/2$ is perfectly stable for acoustics and other linear equations \cite{LB_ED, LBMWavesCurved}.
Next, the new values $f_i^{\rm new}$ are transported to the neighboring cells as
\begin{equation}\label{eq:LBM-Advection}
    f_i(\vec x+\Delta t \vec v_i)=f_i^{\rm new}(\vec x,t)\quad .
\end{equation}
By performing a Chapman-Enskog expansion it is shown that the macroscopic fields \eqref{eq:MacroscopicFields} satisfy in the continuous limit the following conservative equations \cite{ChopardBook}:
\begin{subequations}\label{eq:LB_continuity}
\begin{equation}
    \pdv{\rho_1}{t} + \div{J_1} = 0\quad,
\end{equation}
\begin{equation}\label{eq:LB_tensor}
    \pdv{\vec J_1}{t} + \div{\Pi^{(0)}} = 0\quad,
\end{equation}
\end{subequations}
with
\begin{equation}\label{eq:Pi_tensor}
    \Pi^{(0)} = \sum_i \vec v_i \otimes \vec v_i f_i^{eq}\quad .
\end{equation}
\begin{figure}[b]
    \centering
    \includegraphics[width=0.45\textwidth]{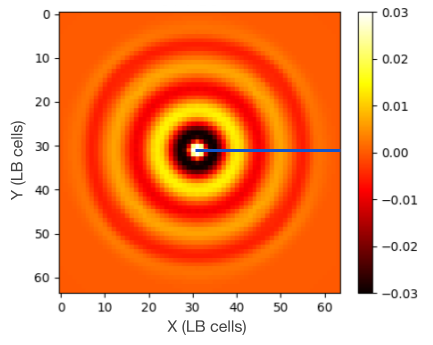}
    \caption{Lattice Boltzmann simulation of the 3D spherical waves produced by a point source after 60 time steps.}
    \label{fig:point_source_map}
\end{figure}
\begin{figure}[b]
    \centering
    \includegraphics[width=0.45\textwidth]{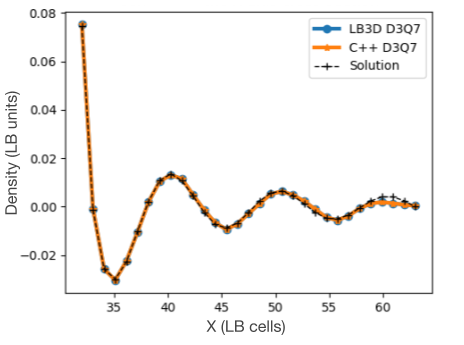}
    \caption{Density profile along the positive $x$-semi-axis and comparison with the analytical solution for the spherical waves of Fig. \ref{fig:point_source_map}.}
    \label{fig:point_source_graph}
\end{figure}
If the equilibrium distribution functions $f_i^{eq}$ are chosen in such a way that $\Pi^{(0)}$ becomes diagonal,
\begin{equation}\label{eq:Pi0_diagonal}
    \Pi^{(0)} = 
    \begin{bmatrix}
        p_1 & 0 \\
        0 & p_1
    \end{bmatrix}\quad\mbox{,}\quad \nabla\cdot\Pi^{(0)}=\nabla p_1\quad,
\end{equation}
taking divergence on both sides of \eqref{eq:LB_tensor} and using \eqref{eq:Pi0_diagonal} and \eqref{eq:LB_tensor} show us that the pressure fulfills 
\begin{equation}\label{eq:LB_waves}
    \pdv[2]{\rho_1}{t} + \nabla^2\rho_1 = 0\quad.
\end{equation}

We can diagonalize $\Pi^{(0)}$ by
choosing a velocity set $\vec v_i$ with weights $w_i$ such that
\begin{eqnarray}\label{eq:velocity_properties}
    \sum_i w_i = 1\quad,\quad \sum_i w_iv_{i\alpha}v_{i\beta} = c_s^2\delta_{\alpha\beta}\\
    \sum_i \omega_i v_{i\alpha} = \sum_i \omega_i v_{i\alpha}v_{i\beta}v_{i\gamma} =0\quad,
\end{eqnarray}
where $v_{i\alpha}$ is the $\alpha-$th component ($\alpha\in x,y,z$) of $\vec v_i$ and $c_s^2$ is a constant. The equilibrium distribution functions
\begin{equation}\label{eq:feq_waves}
    f_i^{eq} = \begin{cases}
        \rho_1(1 + (1/c_s^2) c_0^2(\omega_0-1)) & \text{if}\quad i = 0\\
        (1/c_s^2)\omega_i(c_0^2\rho_1 + \vec v_i \cdot \vec J_1) & \text{if}\quad i \neq 0
    \end{cases}\quad
\end{equation}
make $\Pi^{(0)}$ diagonal (Eq.~\eqref{eq:Pi0_diagonal}), as desired. Here, $c_s$ is the speed of sound and $w_0$ is the weight associated with the null vector $\vec v_0=(0,0)$. Examples are D2Q5, D2Q9 and D3Q19 with $c_s^2=1/3$, and D3Q7, with $c_s^2=1/4$ (Fig.~\ref{fig:velocity_sets}).
The method has been extended to curvilinear cells and has been employed to simulate the normal modes in trumpets and even in the human Cochlea \cite{LBMWavesCurved}. 

Boundary conditions, like a vibrating wall at $x_0=0$, can be set, first, by computing the equilibrium functions $f_i^{eq}$ with the desired value $\rho_1(x_0=0,y,z) = \rho_a\sin(\omega t)$ for the density (whereas $\vec J_1$ is computed from Eq.~\eqref{eq:MacroscopicFields}, as usual) and, next, by overwriting $f_i^{\rm new}$ with that equilibrium value. This operation, which we call \textit{imposing fields}, is performed after collision and before advection, i.e.
\begin{equation}\label{eq:imposingfields}
    f_i^{\rm new}(\vec x,t) = f_i^{eq}\left(\rho_1(x_0=0,y,z),\vec J_1(\vec x,t)\right)\quad .
\end{equation}
The same procedure is employed to set the initial conditions.
Other boundary conditions, like partially absorbing walls at $x_n=Lx-1$ with $Lx$ the length of the simulation domain in $x$ direction, are set by performing a bounce-back step (instead of the usual collision),
\begin{equation}\label{eq:bounce-back}
    f_i(\vec x_n, t_{n+1}) = \Phi f_{j}(\vec x_n, t_n)\quad,
\end{equation}
where $j$ is the index satisfying $\vec v_{j} = -\vec v_i$ 
and $\Phi\in[0,1)$ is a damping factor, which we set as $\Phi=0.99$.
Finally, the results are plotted from the values $f_i^{\rm new}$ after imposing fields and before advection.

\begin{figure}
    \centering
    \includegraphics[width=0.45\textwidth]{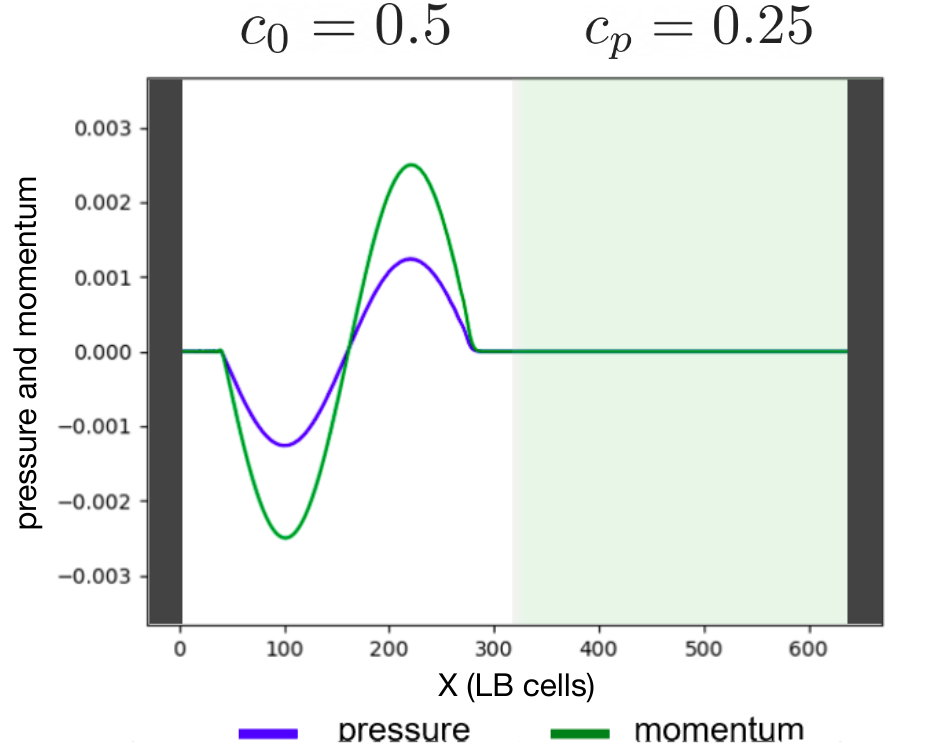}
    \caption{Incident waves in LBM.}
    \label{fig:interface_init}
\end{figure}

\begin{figure}[b]
    \centering
    \includegraphics[width=0.35\textwidth]{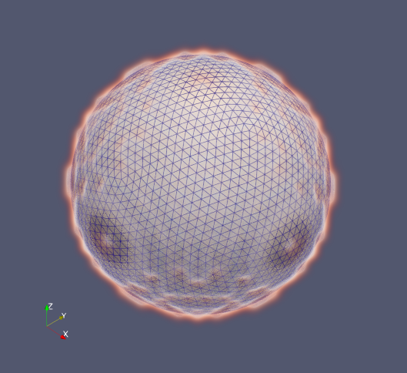}
    \caption{Three-dimensional density map and mesh for the immersed sphere in \textit{3nskog}.}
    \label{fig:mesh}
\end{figure}

The described LBM for waves was implemented in 2D on a self-developed C++ code (available in Github \cite{MyCode}), while for 3D simulations, the software \textit{3nskog} \cite{LB3D} was modified to include this LB model as an additional feature. A first numerical test to validate the code was to simulate the density field in 3D produced by an oscillating point source at the origin, described by the inhomogeneous wave equation
\begin{equation}\label{eq:waves_with_pointSource}
    \nabla^2\rho - \frac{1}{c^2}\frac{\partial^2\rho}{\partial t^2} = \delta(\vec x - \vec x')\rho_0\sin(\omega t)\quad,
\end{equation}
with $\omega$ the angular frequency. The solution is found through Green's functions \cite[Sec.~6.4]{Jackson} and takes the form
\begin{equation}\label{eq:point_solution}
    \rho(\vec x,t) = \frac{\rho_0\sin(\omega t - kr)}{4\pi k r}\quad,
\end{equation}
where $r = |\vec x - \vec x'|$ is the distance from the source position to the measurement point $\vec x$. Because the source term of \eqref{eq:waves_with_pointSource} is proportional to $\sin(\omega t)$, the initial condition $\rho_1(x,t=0) = 0$ everywhere coincides with the source and will not cause numerical instabilities. Figs.~\ref{fig:point_source_map} and \ref{fig:point_source_graph} show a good agreement between the analytical solution \eqref{eq:point_solution} and the 3D implementation and on a cubic domain of $64\times 64\times 64$ lattice cells. 

\begin{figure}
    \centering
    \includegraphics[width=0.45\textwidth]{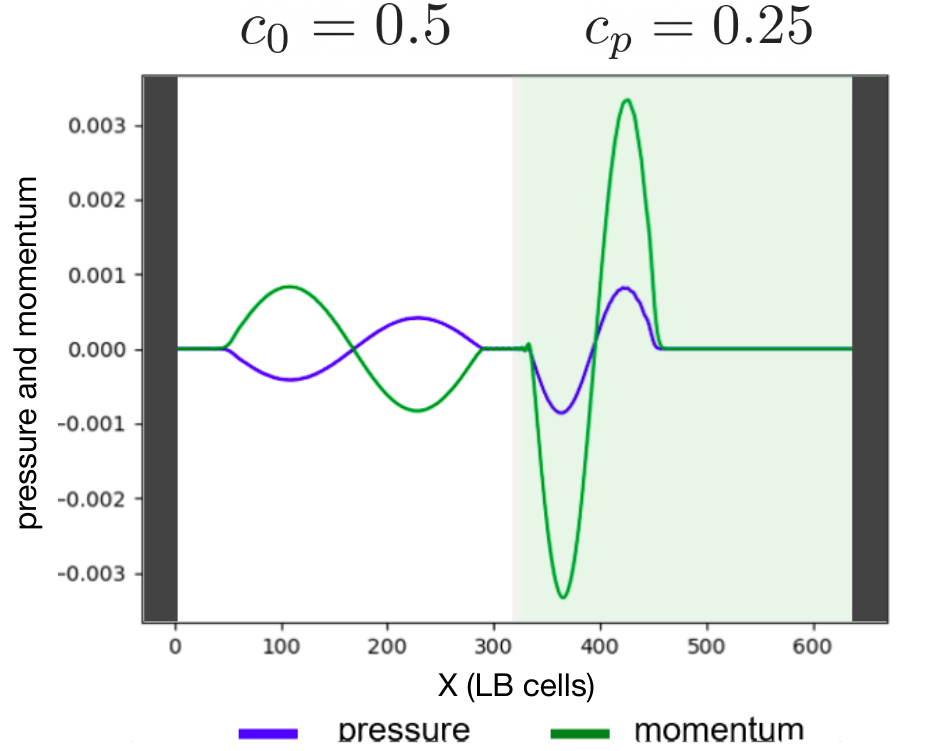}
    \caption{Reflected and transmitted waves in LBM.}
    \label{fig:interface_final}
\end{figure}


A relevant test for our application is the simulation of an interface between two media with different sound speeds $c_0$ and $c_p$
The interface is modeled just by setting a smooth change in the speed of sound along the $x$ axis (to avoid numerical instabilities), 
\begin{equation}\label{eq:c_plane_int}
    c(x) = \frac{c_p - c_0}{2}\tanh{\left(\frac{x-L_x/2}{w}\right)} + \frac{c_p + c_0}{2}\quad,
\end{equation}
with $c_0 = 0.5$ and $c_p = 0.25$, as shown in Figs.~\ref{fig:interface_init} and \ref{fig:interface_final}. This kind of boundary condition generates a typical refraction of waves entering in two media, which is sufficient to reproduce the Gor'kov force.
By setting a traveling pulse as the initial condition, one expects that both pressure and the normal velocity must be continuous at the interface, i.e. that $p_{1,I} + p_{1,R} = p_{1,T}$ and $u_{1x,I} + u_{1x,R} = u_{1x,T}$, with $u_{1x}$ the $x$ component of the velocity and the subindexes $I$,$R$ and $T$ representing the incident, reflected and transmitted values at the interface, respectively. By simulating the interface in a domain of $Lx=$ cells, we obtain that those continuity conditions are fulfilled with 1\% accuracy.

\section{Wave-particle interaction and interpolation}\label{sec:implementation}

To describe the wave-particle interaction in our LBM scheme it is useful to write the acoustic radiation force \eqref{eq:arf_integral} in terms of the quantities $\vec J_1=\rho_0\vec u_1$ and $\rho_1=p_1/c_0^2$ that are computed by the LBM for waves, that is
\begin{equation}\label{eq:arf_integral_LBM}
    \langle F_{i} \rangle = - \frac{1}{\rho_0}\oint \left\langle\left(-\frac{J_1^2}{2}+\frac{\rho_1^2c_0^2}{2}\right)\delta_{ij}+J_{1i}J_{1j}\right\rangle dS_j \quad.
\end{equation}
This expression seems to be just a function of the density $\rho_0$ and the speed of sound $c_0$ in the fluid, but the values $\vec J_1$ and $\rho_1$ depend on the way the interface is modeled. The boundary conditions for the acoustic wave at the interface are that the pressure $p_1$ and the normal velocity $\vec u_1\cdot\hat n$ (with $\hat n$ a vector normal to the surface) must be continuous. Those conditions are equivalent to
\begin{equation}\label{eq:TrueBoundaryConditions}
    \rho_1^{(0)}c_0^2=\rho_1^{(p)}c_p^2
    \quad,\quad
    \frac{\vec J_1^{(0)}\cdot\hat n}{\rho_0^{(0)}} = \frac{\vec J_1^{(p)}\cdot\hat n}{\rho_0^{(p)}} \quad,
\end{equation}
where the superscripts $(0)$ and $(p)$ identify the values of the fields $\vec J_1$ and $\rho_1$ at the fluid side and the object side, respectively.

\begin{figure}
    \centering
    \includegraphics[width=0.35\textwidth]{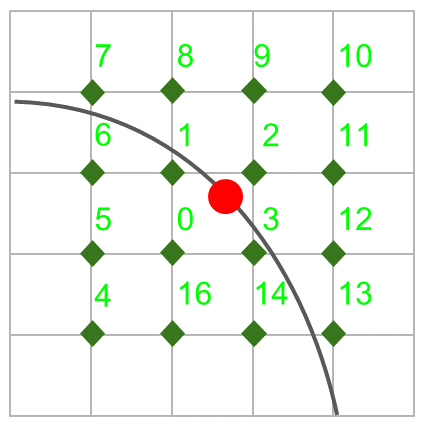}
    \caption{Kernel interpolation scheme.}
    \label{fig:interpolation}
\end{figure}

For the sake of ease, we choose a much simpler approach, assuming that mean densities on both media are equal ($\rho_0^{(0)}=\rho_0^{(p)}$) and that only the speed of sound changes, as
\begin{equation}\label{eq:sound_speed_object}
    c(\vec r) = c_0 - \frac{c_0 - c_p}{2}\left(1+\tanh{\left(\frac{|\vec r-\vec r_0|^2 - R_p^2}{d}\right)}\right)\quad,
\end{equation}
where $\vec r = (x,y,z)$, $\vec r' = (x_0, y_0, z_0)$ and $d$ is the ``square thickness'' of the interface, which was set as $d=4$ for our simulations. In Fig.~\ref{fig:mesh} a three-dimensional density map of function \eqref{eq:sound_speed_object} is illustrated.
This approach is enough to verify that our procedure reproduces Gor'kov and Wei theoretical expressions for the acoustic radiation force on a sphere (3D) and a cylinder (2D), as we will see in the following Section.

The acoustic radiation force \eqref{eq:arf_integral_LBM} is computed by dividing the object's boundary into $N$ surface elements (which are segments in 2D and triangles in 3D) and by computing the normal vector $\hat n$ and the fields $\rho_1$ and $\vec J_1$ at the center $\vec R_k$ of each element, but those points do not usually coincide with the nodes $\vec R^{\text{LB}}$ of the LB mesh (Fig.~\ref{fig:interpolation}). To calculate the fields at $\vec R_k$ we use kernel interpolation \cite{INTERPOLATION,Roma1999}, expressing those fields as a weighted sum on the neighboring cells,
\begin{align}\label{results.mesh.eq:InterpolatedFields}
    \rho_{\text{I}}(\vec R_k)&=\sum_l \rho_1(\vec R_l^{\text{LB}})\delta(|\vec R_l^{\text{LB}}-\vec R_k|)
    \quad,\\
   \vec J_{\text{I}}(\vec R_k)&=\sum_l \vec J_1(\vec R_l^{\text{LB}})\delta(|\vec R_l^{\text{LB}}-\vec R_k|)\quad,
\end{align}
with weights depending on the distance $r=|\vec R_l^{\text{LB}}-\vec R_k|$ through the kernel function \cite{Favier2014} 
\begin{equation}
    \delta(r) = 
    \begin{cases}
        \frac{1}{6}\left(5-3|r|-\sqrt{-3(1-|r|)^2+1}\right) & |r|\in\left[\frac{1}{2},\frac{3}{2}\right]\\
        \frac{1}{3}\left(1+\sqrt{-3r^2+1}\right) & |r|\leq\frac{1}{2}\\
        0 & \text{elsewhere}
    \end{cases}\quad .
\end{equation}

\begin{table}
\centering
\begin{tabular}{|c|c|}
\hline
\textbf{Parameter} & \textbf{Value} \\ \hline
$c_p$              & 0.25           \\ \hline
$c_0$              & 0.24           \\ \hline
$p_0$              & 1.00           \\ \hline
$\lambda$          & 500            \\ \hline
$R_p$              & 10             \\ \hline
$kR_p$             & 0.125         \\ \hline
$x$              & $\lambda/8$           \\ \hline
\end{tabular}
\caption{Default fixed values for the relevant quantities of the acoustic radiation force.}
\label{tab:default_values}
\end{table}

\section{Results and discussion}\label{sec:results}

\begin{figure}[b]
    \centering
    \includegraphics[width=0.5\textwidth]{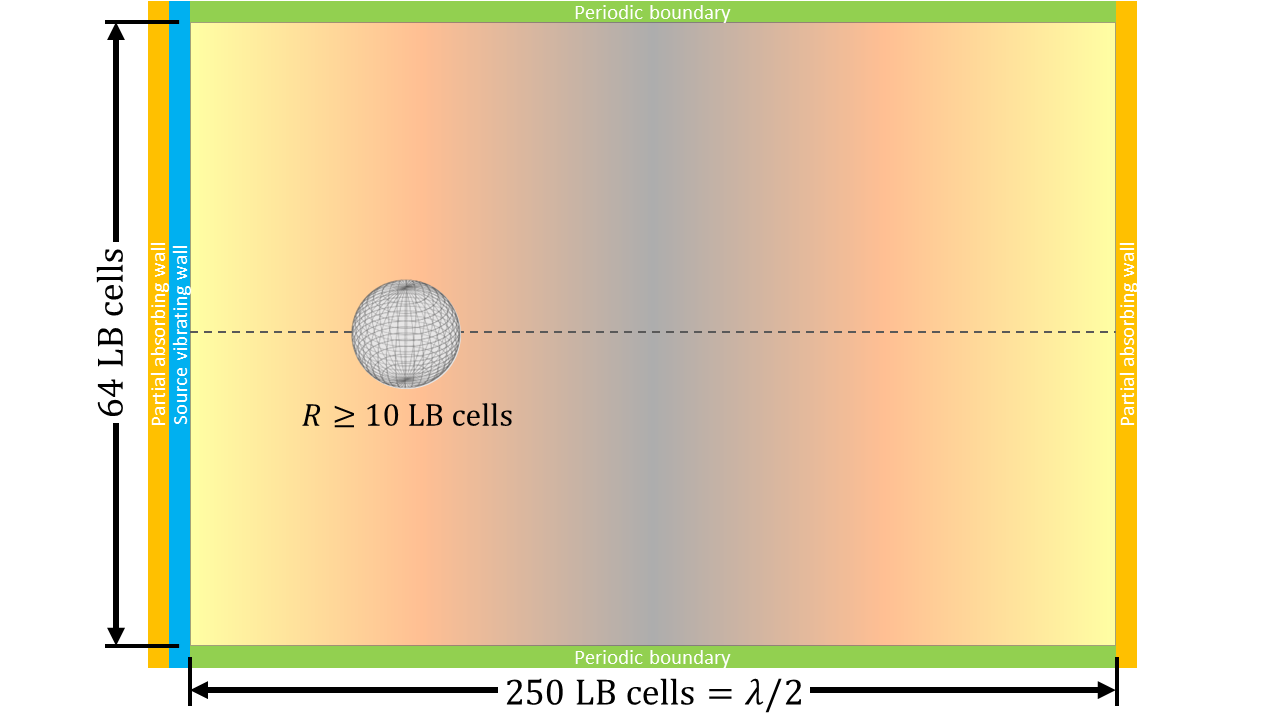}
    \caption{Simulation's domain and setup.}
    \label{fig:setup}
\end{figure}

\begin{figure}
    \centering
    \includegraphics[width=0.5\textwidth]{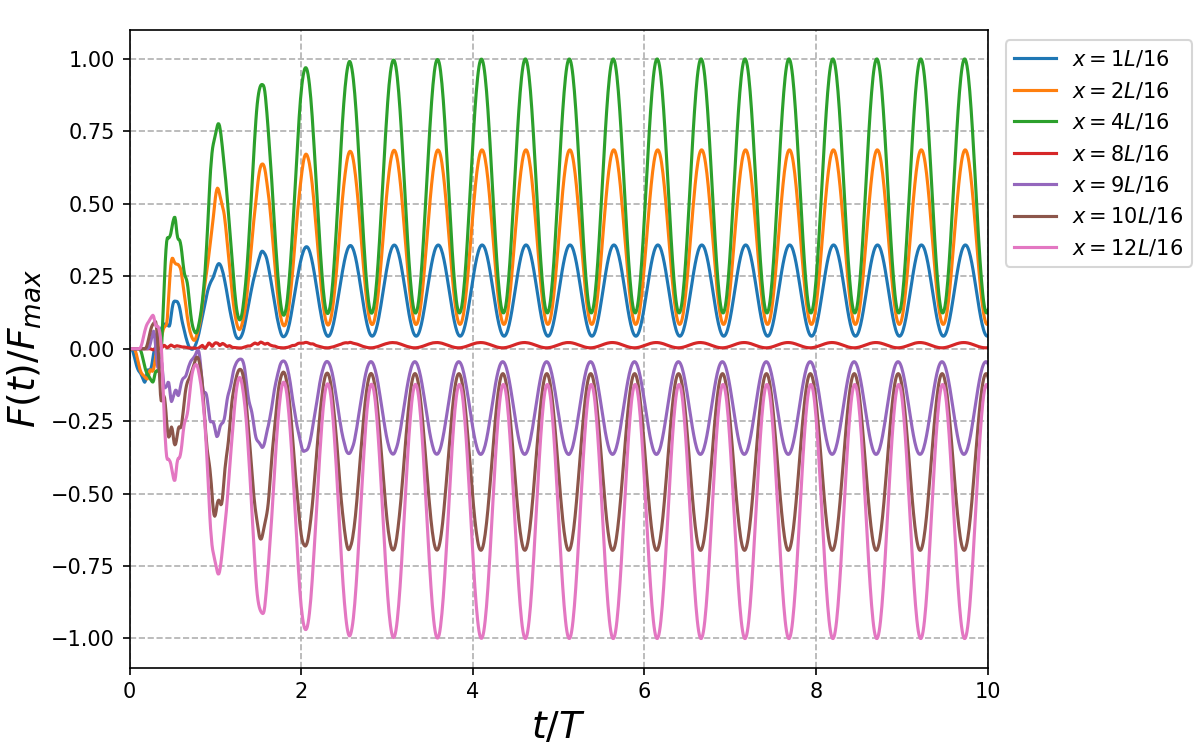}
    \caption{Different curves of the time-dependent force varying the object's axial position.}
    \label{fig:F_t_pos}
\end{figure}

To compute the acoustic radiation force on a sphere and a cylinder and to compare the results with the Gor'kov and Wei solutions, respectively, the simulation domain is a box with $L_y=L_z=64$ cells along $y$ and $z$ axis in 3D and $L_y=64$ cells in 2D. The length $Lx$ along the $x$ axis is chosen such that $Lx=\lambda/2+3$, with $100\le \lambda \le 2560$ the wavelength in cell units (Fig.~\ref{fig:setup}). The object, a sphere or a cylinder of radius $R_p$, is placed with its center at $\vec r_0=(x_0,L_y/2,L_z/2)$ and the speed of sound was chosen at every cell by using \eqref{eq:sound_speed_object}. The default values for the mean density $\rho_0$, the wavelenght $\lambda$, the object's radius $R_p$, the speed of sound in the fluid $c_0$ and in the object, $c_p$, are listed in Table~\ref{tab:default_values}. The default value for $kR_p$ in all test where neither $k$ nor $R_p$ were varied is $0.125$. This value is small enough to compare with Gor'kov solutions. The left wall, chosen as source, is built with two extra sheets: one partial absorbing wall at $x=0$ and one source of plane waves at $x=1$, forcing the density to oscillate as $\rho_a\sin(\omega t)$ (Eq.~\eqref{eq:imposingfields}). The right wall, at $x=L_x+2$ (i.e. the last cells in the $x$ axis) is chosen as a partially absorbing reflector (Eq.~\eqref{eq:bounce-back}). In addition, periodic boundary conditions are set on $y$ and $z$ directions.
The result is a standing wave with a single nodal plane at $x=L_x/2+1$. 

\begin{figure}[b]
    \centering
    \includegraphics[width=0.48\textwidth]{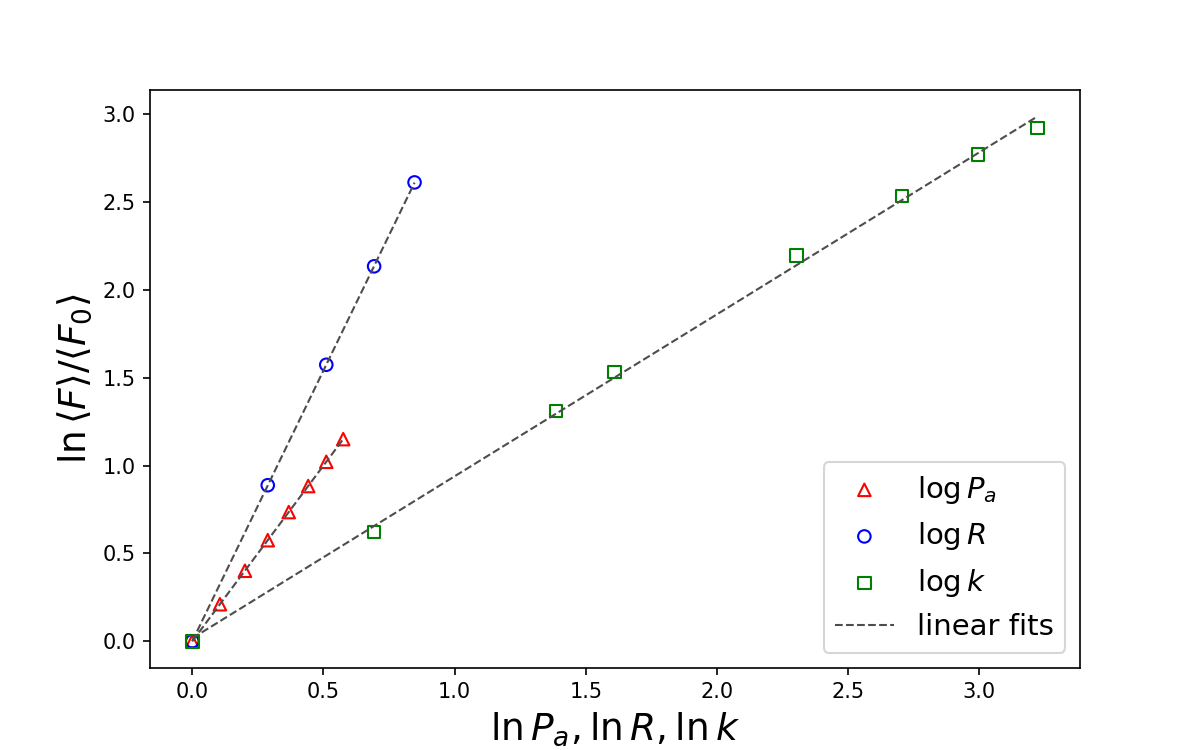}
    \caption{Acoustic radiation force of the sphere (3D) for several pressure amplitudes $P_a$, sphere's radii $R_p$ and wave numbers $k$. For $k$ curve $kR_p$ were from $0.02$ to $0.62$, while for $R_p$ from $0.04$ to $0.10$.}
    \label{fig:amp_R_k_3D}
\end{figure}

\begin{figure}
    \centering
    \includegraphics[width=0.5\textwidth]{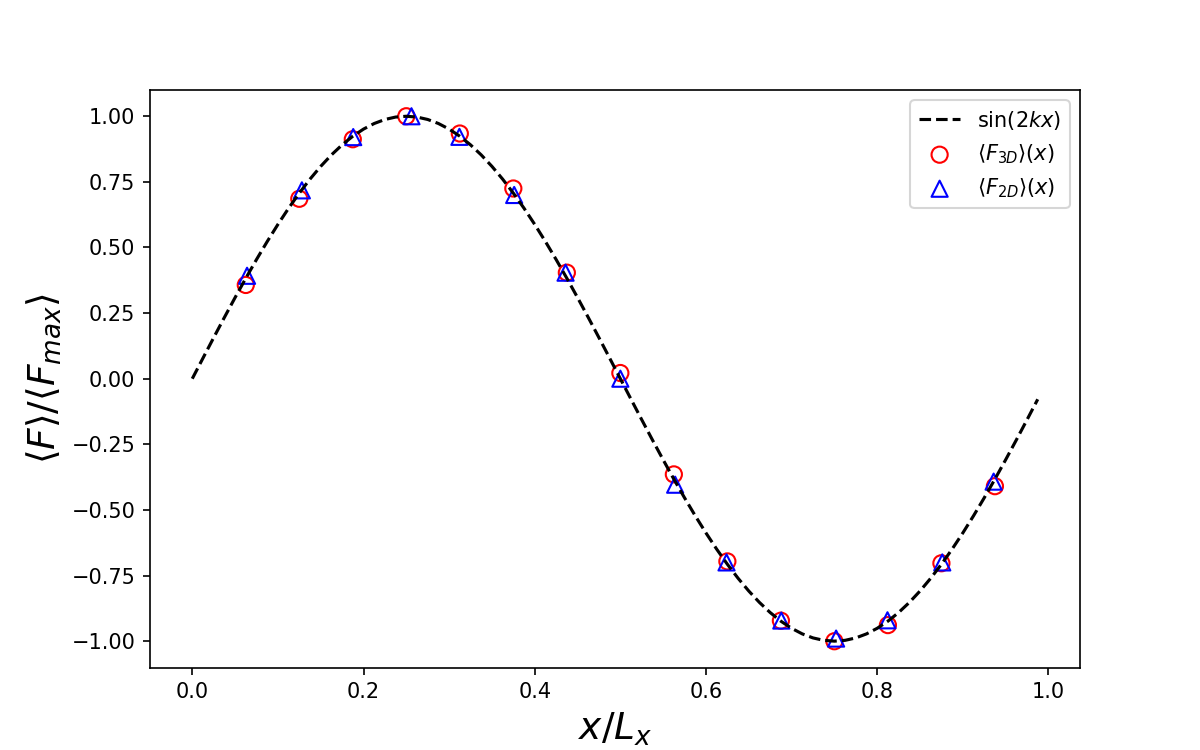}
    \caption{Acoustic radiation forces in 2D and 3D along the $x$ axis showing a $\sin 2kx$ dependence. }
    \label{fig:position2D3D}
\end{figure}

Each simulation runs for more than $10T$ time steps, with $T=\lambda/c_0$ the period of acoustic oscillations.  The acoustic radiation force is measured at every time step using \eqref{eq:arf_integral_LBM}, giving us a curve of $F_x(t)$ vs. time that stabilizes after around ten oscillations. Once it stabilizes, $\left< F_x \right>$ is measured as that signal's average (Fig.~\ref{fig:F_t_pos}).

\begin{figure}[b]
    \centering
    \includegraphics[width=0.48\textwidth]{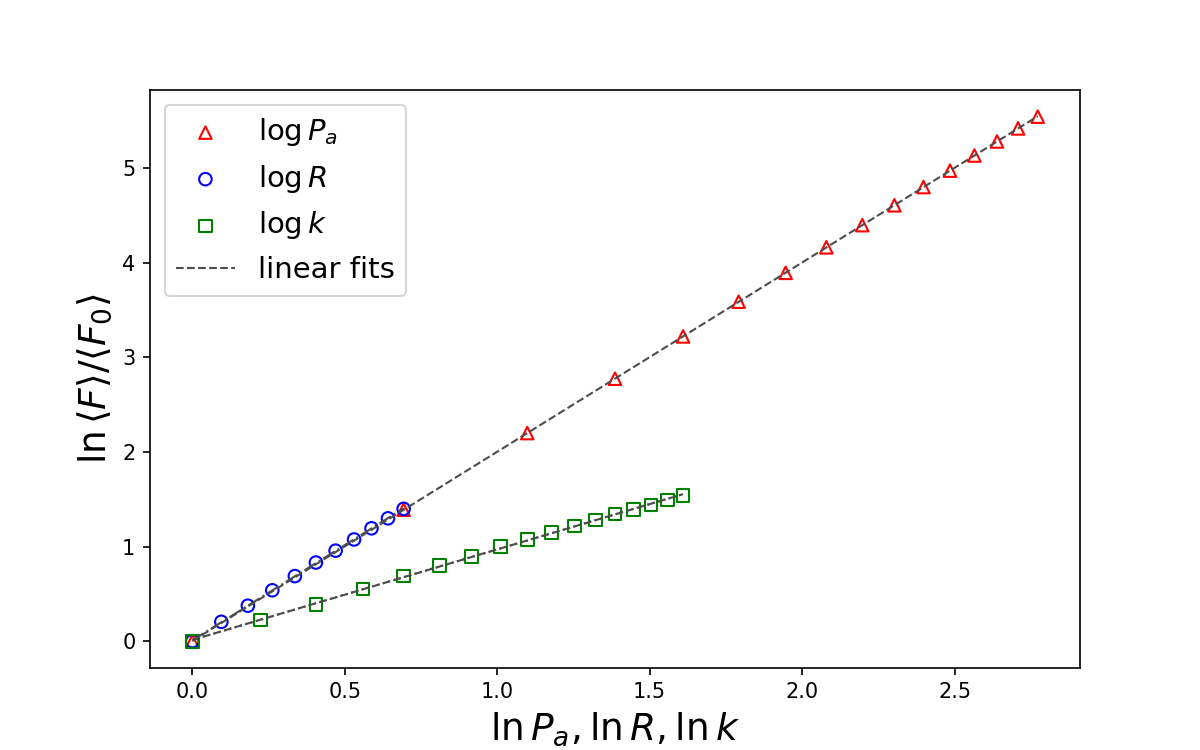}
    \caption{Acoustic radiation force of the cylinder (2D) for several pressure amplitudes $P_a$, cylinder's radii $R_p$ and wave numbers $k$. For $k$ curve $kR_p$ were from $0.02$ to $0.62$, while for $R_p$ from $0.04$ to $0.10$.}
    \label{fig:amp_R_k_2D}
\end{figure}

\begin{figure}
    \centering
    \includegraphics[width=0.48\textwidth]{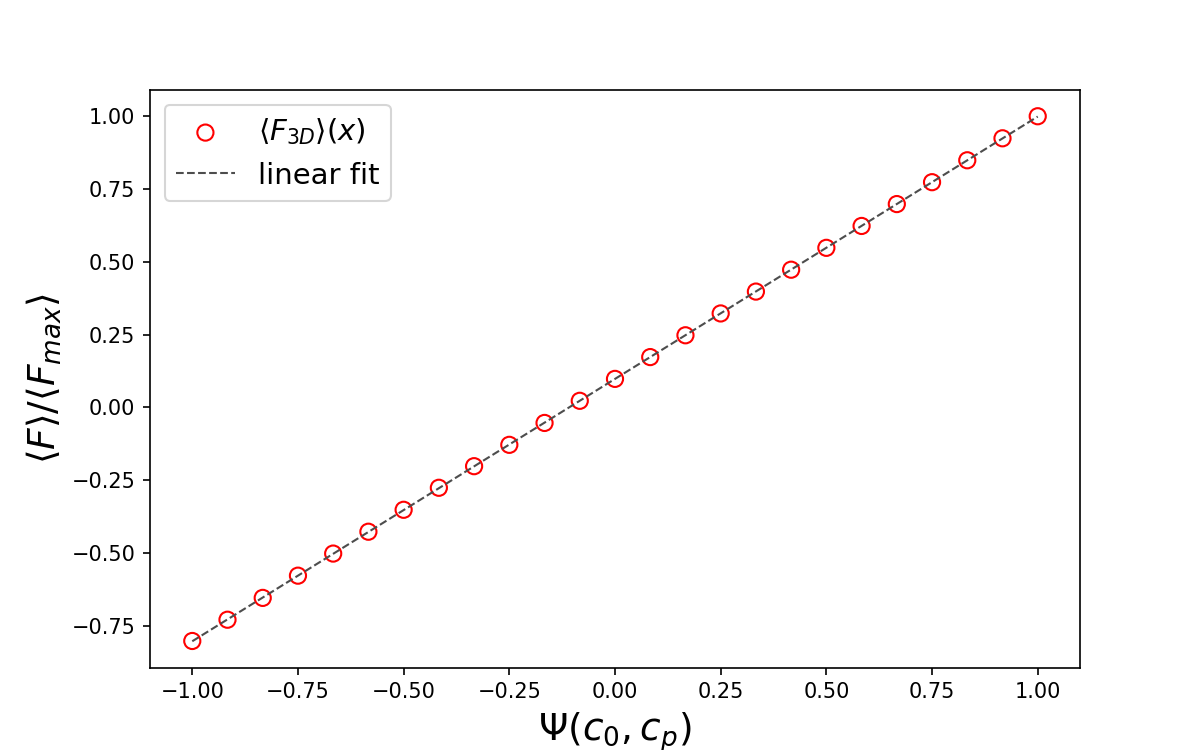}
    \caption{Acoustic radiation force of the sphere (3D) for several values of the contrast factor $\Psi$ (Eq. \eqref{eq:speeds_factor}) with $kR_p=0.125$.}
    \label{fig:speed3D}
\end{figure}

\begin{table}[b]
\begin{tabular}{|c|c|c|c|}
\hline
\textbf{Parameter} & \textbf{Expected} & \textbf{Measured} & \textbf{Rel. Err. (\%)} \\ \hline
\textbf{$P_a$}     & $2.0000$          & $1.9999$          & $10^{-13}$              \\ \hline
\textbf{$k$}       & $1.0000$          & $0.9222$          & $7.7783$                \\ \hline
\textbf{$R_p$}     & $3.0000$          & $3.0814$          & $2.7119$                \\ \hline
\end{tabular}
\caption{Power law exponents comparison in 3D.}
\label{tab:results_power_3D}
\end{table}



The measurements are compared with the theoretical expressions \eqref{eq:ARF_Gorkov_3D} and \eqref{eq:ARF_Gorkov_2D} for the sphere and the cylinder, respectively, with equal mean densities for the fluid and the object, that we rewrote as
\begin{subequations}\label{eq:expected_forces}
    \begin{equation}\label{eq:expected_arf_3D}
    \langle F_x\rangle_{\text{3D}} = -\frac{\pi R_p^3P_a^2k}{3\rho_0}\Psi(c_0, c_p)\sin 2kx \mbox{ ,} 
    \end{equation}
    \begin{equation}\label{eq:expected_arf_2D}
    \langle F_x\rangle_{\text{2D}} = -\frac{\pi R_p^2P_a^2k}{4\rho_0}\Psi(c_0, c_p)\sin 2kx \mbox{ .} 
    \end{equation}
\end{subequations}
Hereby we have defined the contrast factor $\Psi(c_0, c_p)$ as
\begin{equation}\label{eq:speeds_factor}
    \Psi(c_0,c_p) = \frac{1}{c_0^2} - \frac{1}{c_p^2}\quad.
\end{equation}
Simulations were performed by varying one at a time the physical parameters the force depends on, whereas the other parameters remain fixed to default values (Tab.~\ref{tab:default_values}). The varied parameters are the axial position of the object $x$, the pressure acoustic amplitude $P_a$, the radius of the object $R_p$ and the wave number $k=2\pi/\lambda$. The contrast factor $\Psi(c_0,c_p)$ is also properly changed to find out its linear relationship with the force. 

Figure \ref{fig:position2D3D} shows that the simulated force follows a behavior proportional to $\sin(2kx)$ for both 3D and 2D cases, being the force zero when the object is at $x= 8L/16 = \lambda/4$ where the node is (see Fig.~\ref{fig:F_t_pos}). When $\Psi(c_0, c_p)>0$, this is a confining force at the nodal plane. 
Concerning $R_p$, $k$ and $P_a$, we can see from \eqref{eq:expected_forces} that a power law relationship with the force is expected for each one of them. Figures \ref{fig:amp_R_k_3D} and \ref{fig:amp_R_k_2D} show that this is the case for both the sphere and the cylinder, with  correlation coefficients close to $0.99$. The measured exponents for those power laws show an excellent agreement with the expected values.(Tabs.~\ref{tab:results_power_3D} and \ref{tab:results_power_2D}). As noticed in figure \ref{fig:amp_R_k_3D} the linear relation for $k$ has some deviations in the last points from the fitted curve, and the measured parameters $k$ and $R_p$ differs in $8\%$ and $3\%$ from their theoretical values (Table \ref{tab:results_power_3D}). In Fig. \ref{fig:amp_R_k_3D} the $kR_p$ values for the  $\langle F\rangle$ vs. $k$ curve run from $0.02$ to $0.62$, and from $0.04$ to $0.10$ for the $\langle F\rangle$ vs. $R_p$ curve. Since the theoretical results we are compared with are based on the assumption $kR_p\ll 1$, one could expect some deviations of the theory from the computational results. Better theoretical models, as those proposed by Marston in \cite{n7_Marston2017}, include higher order expansions in $kR_p$. A comparison with that high order expressions can be a future continuation of this work. The influence of resolution has not been studied in this work, but we find plausible to see an reduction of the relatives errors reported in tables \ref{tab:results_power_3D} and \ref{tab:results_power_2D}.

\begin{figure}
    \centering
    \includegraphics[width=0.48\textwidth]{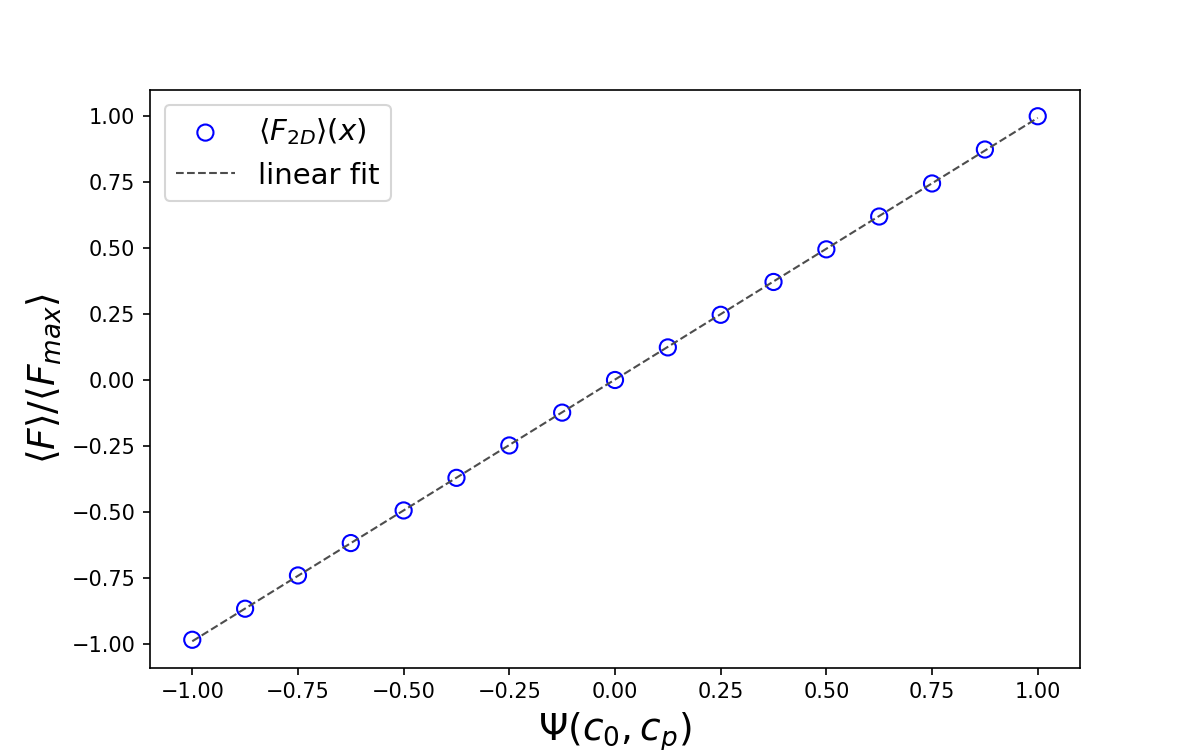}
    \caption{Acoustic radiation force of the cylinder (2D) for several values of the contrast factor $\Psi$ (Eq. \eqref{eq:speeds_factor}) with $kR_p=0.125$.}
    \label{fig:speed2D}
\end{figure}

Regarding the contrast factor, we expect a linear relationship between $\left< F_x \right>$ and $\Psi(c_0,c_p)$ (Eq.~\eqref{eq:expected_forces}). For this test, the speed of sound for the fluid was set to $c_0 = 0.25$, whereas $c_p$ was adjusted so that the factor $\Psi$ changes uniformly. Figures \ref{fig:speed3D} and \ref{fig:speed2D} confirms that linear behavior for both the sphere and the cylinder, with $R^2 = 0.994$ and $R^2 = 0.999$ for the 3D and 2D cases, respectively. 

Summarizing, our simulations show that the proposed method to compute the acoustic radiation force reproduces with great precision both the theoretical expressions by Gor'kov and Wei.

\section{Conclusions}

We employed a lattice Boltzmann model that reproduces the wave equation to compute the acoustic radiation force on a compressible object immersed in an inviscid fluid. Instead of simulating the full Navier-Stokes equations (NSE), our proposal relies on the solution of the wave equation from which, via an interpolation scheme, the first order perturbations in the velocity $\vec u_1$ and the pressure $p_1$ are computed on the object's surface and, from them, the force.
We tested our novel numerical approach by computing the acoustic radiation force on a cylinder (in 2D) and on a sphere (in 3D) produced by a standing wave with a single nodal plane. All tests were done in the density matched case.
The method reproduces with excellent accuracy the theoretical predictions by Gor'kov~\cite{Gorkov1962} and Wei et al.~\cite{Wei2004} for the sphere and the cylinder, respectively, in all studied cases.
The procedure to compute the acoustic radiation force from the fields $p_1$ and $\vec u_1$ is general, and can be performed with any numerical method for solving the wave equation.

\begin{table}[b]
\begin{tabular}{|c|c|c|c|}
\hline
\textbf{Parameter} & \textbf{Expected} & \textbf{Measured} & \textbf{Rel. Err. (\%)} \\ \hline
\textbf{$P_a$}     & $2.0000$          & $1.9994$          & $0.0302$                \\ \hline
\textbf{$k$}       & $1.0000$          & $0.9569$          & $4.3108$                \\ \hline
\textbf{$R_p$}     & $2.0000$          & $2.0127$          & $0.6373$                \\ \hline
\end{tabular}
\caption{Power law exponents comparison in 2D.}
\label{tab:results_power_2D}
\end{table}


One of the advantages of our numerical approach is that it 
computes $p_1$ and $\vec u_1$, with second order accuracy, and hence it needs fewer cells to compute the acoustic radiation force than other methods simulating the full Navier-Stokes equations. Actually, widely used LBMs for fluids compute with second order accuracy the zeroth order density and velocity fields and, with only first order accuracy the perturbative fields $p_1$ and $\vec u_1$. Accordingly, our method obtains excellent results by using grids of just $250\times 64$ cells for the 2D case and $250\times 64\times 64$ for the 3D one, whereas a previous work using a LBM for the full NSE \cite{cosgrove2004} requires $600\times 600$ cells to obtain reliable results in 2D. Our method is also very fast. When implemented in C++ it takes up to $3$ minutes on a laptop with a single Intel Core-i5 processor to compute the acoustic radiation force on a cylinder with the simulation domain described above. Similarly, when implemented in our 3D solver \textit{3nskog} it takes no longer than $30$ minutes by running on four processors to compute the acoustic radiation force on a sphere.

The values of $p_1$ and $\vec u_1$ on the object's surface and the acoustic radiation force computed from them rely on the boundary conditions employed to model the object itself. For simplicity, in our tests we used just a smooth change in the speed of sound when crossing from the fluid to the object, assuming that the mean densities in both media are equal (which is the density-matched case). On the one hand, such an approximation can be enough for nearly buoyant particles. On the other hand, this strategy assures that both the pressure and the normal velocity are continuous at the interface. The actual boundary conditions (Eq.~\eqref{eq:TrueBoundaryConditions}) relate both the momenta and the densities on both sides of the interface. Finding the way to implement those more general boundary conditions in our LBM will be an interesting subject for future work.

The proposed approach can be used to compute the acoustic radiation force on an object of any shape in 2D or 3D, like microbots or self-assembled microswimmers. Moreover, it can be coupled with standard numerical integration algorithms 
to compute the three-dimensional movement of such objects due to acoustic radiation forces and introducing viscous drag forces when needed, according to \cite{m2_danilov2000} (see Supplementary Material \texttt{Movie.gif}).
This opens a broad spectrum of future applications in the acoustic manipulation of many objects. 
In addition, being able to have both waves and fluids in the same LBM code allows to simulate multi-physics applications efficiently. For example, the method could be employed for the study of systems like acoustic-driven bubbles or reactions among fluids pushed by micro-particles of solid catalyst driven by acoustic forces, just to mention a few.
All those are interesting subjects for future research.



The present work introduces a fast and accurate numerical procedure to compute the acoustic radiation force on an object immersed in an inviscid fluid. The proposed procedure shows to be a promising tool for the study of the many phenomena in medicine and engineering where that force plays a relevant role.

\section{Acknowledgments}

We acknowledge financial support from the Bavarian University Center for Latin America (BAYLAT). Also, we thank the {\it 3nskog} developer team, in particular Johannes Hielscher and Dr. Othmane Aouane, whose advice and technical support were fundamental to develop the simulation code.

\appendix
\section{Deducting the second order acoustic radiation force}\label{ap:arf_second}
By replacing the second order expansion written in \eqref{eq:expand_2} into the mass (Eq.~\eqref{eq:mass_cons}) and momentum (Eq.~\eqref{eq:euler_vec}) conservation law, using Eq.~\eqref{eq:linear_NSE_eqs} and taking the second order terms only, we obtain
\begin{subequations}\label{eq:nonlinear_NSE_eqs}
\begin{equation}\label{eq:nonlinear_mass_conservation}
    \frac{\partial\rho_2}{\partial t} + \rho_0\nabla\cdot\vec u_2 + \nabla\cdot(\rho_1\vec u_1) = 0\quad,
\end{equation}    
\begin{equation}\label{eq:nonlinear_momentum_conservation_law}
    \rho_0\frac{\partial\vec u_2}{\partial t} + \rho_1\frac{\partial\vec u_1}{\partial t} + \nabla p_2 + \rho_0(\vec u_1\cdot\nabla)\vec u_1 = 0\quad.
\end{equation}
\end{subequations}
The last term of the left side of Eq.~\eqref{eq:nonlinear_momentum_conservation_law} can be rewritten by using the following mathematical property,
\begin{equation}\label{eq:double_dev_property}
    \frac{1}{2}\nabla(\vec u_1\cdot\vec u_1) = (\vec u_1\cdot\nabla)\vec u_1 + \vec u_1 \times (\nabla \times \vec u_1)\quad,
\end{equation}
where the term $\vec u_1 \times (\nabla \times \vec u_1)=0$, because the flow is irrotational. Then \eqref{eq:nonlinear_momentum_conservation_law} becomes
\begin{equation}
    \rho_0\frac{\partial\vec u_2}{\partial t} + \rho_1\frac{\partial\vec u_1}{\partial t} + \nabla p_2 + \frac{\rho_0}{2}\nabla(u_1^2) = 0\quad.
\end{equation}
By using Eqns.~\eqref{eq:eq_state} and \eqref{eq:linear_momentum_conserv}, the second term of the left hand side may be written as
\begin{equation}
    \rho_0\frac{\partial\vec u_2}{\partial t} - \frac{p_1}{\rho_0 c^2}\nabla p_1 + \nabla p_2 + \frac{\rho_0}{2}\nabla(u_1^2) = 0
\end{equation}
and using the product derivative property for gradients, we end up with
\begin{equation}
    \rho_0\frac{\partial\vec u_2}{\partial t} + \nabla p_2 =  \frac{1}{2\rho_0 c_0^2}\nabla(p_1^2) - \frac{\rho_0}{2}\nabla(u_1^2)\quad.
\end{equation}

Because we are interested in writing the total velocity and pressure fields in terms of only first order terms, let us add Eq.~\eqref{eq:linear_momentum_conserv} as a null term such that, by using \ref{eq:expand_p_2} and \eqref{eq:expand_u_2} we have
\begin{align}
    \rho_0\frac{\partial\vec u_2}{\partial t} + \rho_0\frac{\partial\vec u_1}{\partial t} + \nabla p_2 + \nabla p_1 &=  \frac{1}{2\rho_0 c_0^2}\nabla(p_1^2) - \frac{\rho_0}{2}\nabla(u_1^2)\nonumber\,,\\
    \rho_0\frac{\partial\vec u}{\partial t} + \nabla (P - p_0) &= \nabla\left(\frac{1}{2\rho_0 c^2}p_1^2 - \frac{\rho_0}{2}u_1^2\right)\quad.
\end{align}
By using \eqref{eq:grad_phi_u1_p1} we end up with a total non-static pressure \begin{equation}\label{eq:euler_king_grad_ap}
    \nabla(P - p_0) = \nabla\left(\frac{p_1^2}{2\rho_0 c_0^2} - \frac{\rho_0}{2}u_1^2 - \rho_0\frac{\partial\phi}{\partial t}\right)
\end{equation}
written also in \eqref{eq:euler_king_grad}.

\section{Incident, scattered and interference terms}\label{ap:in_sc_force}

With the velocity potential divided into an incident and a scattered field, the average force Eq.~\eqref{eq:ARF_general} has three contributions. The first one, due to $\phi_{\text{in}}$ only, 
\begin{align}\label{eq:force_in_in}
    \langle F_{i,\textsc{in}} \rangle &=-\rho_0\oint\bigg\langle\left(-\frac{\rho_0}{2}|\nabla\phi_{\text{in}}|^2 + \frac{\rho_0}{2c_0^2}\left[\frac{\partial\phi_{\text{in}}}{\partial t}\right]^2\right)\delta_{ij} \nonumber\\
    &+ \rho_0\partial_i\phi_{\text{in}}\partial_j\phi_{\text{in}}\bigg\rangle dS_i
\end{align}
should be zero because the incident field (which is the solution in the absence of the object) does not receive any physical effect from the particle. In the case of plane waves, the incident field $\phi_{\text{in}} = \phi_0 \cos(\vec k\cdot \vec r - \omega t)$ is spatially homogeneous, implying a symmetry over the surface, and the closed integral will yield zero~\cite[~p.79]{Manneberg2009}\cite[~p.]{Bruus2012_07}.
The other two contributions containing information about the scattered wave are
\begin{align}\label{eq:force_sc_sc}
    \langle F_{i,\textsc{sc}} \rangle &= -\rho_0\oint\bigg\langle\left(-\frac{\rho_0}{2}|\nabla\phi_{\text{sc}}|^2 + \frac{\rho_0}{2c_0^2}\left[\frac{\partial\phi_{\text{sc}}}{\partial t}\right]^2\right)\delta_{ij}\nonumber\\ 
    &+\rho_0\partial_i\phi_{\text{sc}}\partial_j\phi_{\text{sc}}\bigg\rangle dS_i
\end{align}
and
\begin{align}\label{eq:force_in_sc}
    \langle F_{i,\textsc{in-sc}} \rangle &=-\rho_0\oint\bigg\langle\bigg(-\rho_0\nabla\phi_{\text{in}}\cdot\nabla\phi_{\text{sc}} + \nonumber\\
    &\frac{\rho_0}{c_0^2}\bigg[\frac{\partial\phi_{\text{in}}}{\partial t}\bigg]\bigg[\frac{\partial\phi_{\text{sc}}}{\partial t}\bigg]\bigg)\delta_{ij} +\nonumber\\ &\rho_0\partial_i\phi_{\text{in}}\partial_j\phi_{\text{sc}} + \rho_0\partial_i\phi_{\text{sc}}\partial_j\phi_{\text{in}}\bigg\rangle dS_i\quad.
\end{align}
The contribution $\langle F_{i,\textsc{sc}} \rangle$ is much smaller than the interference term $\langle F_{i,\textsc{in-sc}} \rangle$ because the scattering cross-section of a spherical particle is proportional to $(kR_p)^4$, which is negligible due to $R_p\ll\lambda$, and because the scattered potential field solution is proportional to $R_p^3$, as will be shown later. Thus, the interference term \eqref{eq:force_in_sc} is the most relevant, and that is the one to be developed next~\cite[~p.79]{Manneberg2009}. By using \eqref{eq:grad_phi_u1_p1}
\begin{subequations}\label{eq:phi_fields}
\begin{equation}\label{eq:force_grad_vel}
    \nabla\phi_{in}\cdot\nabla\phi_{sc} = \vec u_{in}\cdot\vec u_{sc}\quad,
\end{equation}    
\begin{equation}\label{eq:force_phi_rho}
    \frac{\rho_0}{c_0^2}\frac{\partial\phi_{in}}{\partial t}\frac{\partial\phi_{sc}}{\partial t} = \frac{c_0^2}{\rho_0}\rho_{in}\rho_{sc}\quad,\text{ and}
\end{equation}    
\begin{equation}
    \partial_i\phi_{\text{in}}\partial_j\phi_{\text{sc}} = u_{\text{in}}^{i}u_{\text{sc}}^{j}
\end{equation}
\end{subequations}
the interference term becomes
\begin{align}\label{eq:force_in_sc2}
    \langle F_{i,\textsc{in-sc}} \rangle &=-\rho_0\oint\bigg\langle\left(-\rho_0\vec u_{\text{in}}\cdot\vec u_{\text{sc}} + \frac{c_0^2}{\rho_0}\rho_{in}\rho_{sc}\right)\delta_{ij}\nonumber\\
    &+ \rho_0 u_{\text{in}}^{i}u_{\text{sc}}^{j}+ \rho_0u_{\text{sc}}^{i}u_{\text{in}}^{j}\bigg\rangle dS_i\quad.
\end{align}
By using Gauss' theorem, the  surface integral transforms into a volume integral, 
\begin{align}\label{eq:force_in_sc_gauss}
    &\langle F_{i,\textsc{in-sc}} \rangle =-\rho_0\int\bigg\langle\bigg(-\rho_0(\partial_iu_{\text{in}}^{m})u_{\text{sc}}^{m} - \rho_0 (\partial_iu_{\text{sc}}^{m})u_{\text{in}}^{m} +\nonumber\\
    &\frac{c_0^2}{\rho_0}\partial_i\rho_{in}\rho_{sc} + \frac{c_0^2}{\rho_0}\rho_{in}\partial_i\rho_{sc}\bigg)\delta_{ij} + \rho_0(\partial_{i}u_{\text{in}}^{i})u_{\text{sc}}^{j} \nonumber\\
    &+ \rho_0u_{\text{in}}^{i}(\partial_{i}u_{\text{sc}}^{j}) + \rho_0(\partial_{i}u_{\text{sc}}^{i})u_{\text{in}}^{j} + \rho_0u_{\text{sc}}^{i}(\partial_{i}u_{\text{in}}^{j})\bigg\rangle dV\quad.
\end{align}
The first two terms may be rewritten as
\begin{align}
    (\partial_iu_{\text{in}}^{m})u_{\text{sc}}^{m} + (\partial_iu_{\text{sc}}^{m})u_{\text{in}}^{m} &=(\partial_i\partial_m\phi_{\text{in}})u_{\text{sc}}^{m} + (\partial_i\partial_m\phi_{\text{sc}})u_{\text{in}}^{m}\,, \nonumber\\
    &=(\partial_m\partial_i\phi_{\text{in}})u_{\text{sc}}^{m} + (\partial_m\partial_i\phi_{\text{sc}})u_{\text{in}}^{m}\,, \nonumber\\
    &= (\partial_mu_{\text{in}}^{i})u_{\text{sc}}^{m} + (\partial_mu_{\text{sc}}^{i})u_{\text{in}}^{m} 
\end{align}
and the interference term simplifies to
\begin{align}
    \langle F_{i,\textsc{in-sc}} \rangle &=-\rho_0\int\bigg\langle\frac{c_0^2}{\rho_0}\partial_j\rho_{in}\rho_{sc} + \frac{c_0^2}{\rho_0}\rho_{in}\partial_j\rho_{sc} \nonumber\\
    &+ \rho_0(\partial_{i}u_{\text{in}}^{i})u_{\text{sc}}^{j} + \rho_0(\partial_{i}u_{\text{sc}}^{i})u_{\text{in}}^{j}\bigg\rangle dV\quad .
\end{align}

Now, by using \eqref{eq:linear_NSE_eqs} for the incident and scattered fields on all terms but the last one, we obtain
\begin{align}
    &\langle F_{i,\textsc{in-sc}} \rangle =-\rho_0\int\bigg\langle-\frac{\partial u_{\text{in}}^{j}}{\partial t}\rho_{sc} - \frac{\partial u_{\text{sc}}^{j}}{\partial t}\rho_{in} -\frac{\partial\rho_{\text{in}}}{\partial t}u_{\text{sc}}^{j} \nonumber\\    &+\rho_0(\partial_{i}u_{\text{sc}}^{i})u_{\text{in}}^{j}\bigg\rangle dV 
    -\nonumber\\
    &\int\bigg\langle-\frac{\partial u_{\text{in}}^{j}}{\partial t}\rho_{sc} - \frac{\partial}{\partial t}(u_{\text{sc}}^{j}\rho_{in}) + \rho_0(\partial_{i}u_{\text{sc}}^{i})u_{\text{in}}^{j}\bigg\rangle dV \quad.
\end{align}
Since
\begin{equation}
    -\frac{\partial u_{\text{in}}^{j}}{\partial t}\rho_{sc} = -\frac{\partial}{\partial t}(u_{\text{in}}^{j}\rho_{sc}) + u_{\text{in}}^{j}\frac{\partial \rho_{sc}}{\partial t}
\end{equation}
the averaged force simplifies to
\begin{align}
    \langle F_{i,\textsc{in-sc}} \rangle &=-\rho_0\int\bigg\langle-\frac{\partial}{\partial t}(u_{\text{sc}}^{j}\rho_{in} + u_{\text{in}}^{j}\rho_{sc}) +\nonumber\\
    &u_{\text{in}}^{j}\frac{\partial \rho_{sc}}{\partial t} + \rho_0(\partial_{i}u_{\text{sc}}^{i})u_{\text{in}}^{j}\bigg\rangle dV\quad .
\end{align}
Because the time-average of the time derivatives of any periodic function is identically zero, the force simplifies further to
\begin{equation}\label{eq:F_sc_ap}
    \langle \vec F \rangle = - \rho_0\int \left\langle \vec u_{\text{in}}\left(\nabla^2\phi_{\text{sc}}-\frac{1}{c_0^2}\frac{\partial^2\phi_{\text{sc}}}{\partial  t^2}\right)\right\rangle dV\quad.
\end{equation}
Therefore, the time-averaged force on the small sphere can be found if we compute the scattered velocity potential $\phi_{sc}$. 

\section{Integrating the acoustic radiation force to obtain the Gor'kov potential}\label{ap:integrate_gorkov}

With \eqref{eq:a_3D} and \eqref{eq:A_3D} it is now possible to write a particular solution for the scattered velocity potential previously defined in terms of $a(t')$ and $\vec A(t')$ (Eq. \eqref{eq:scattered}). It becomes
\begin{equation}
    \phi_{\text{sc}}(r,t) = -f_1\frac{R_p^3}{3\rho_0 r}\dot{\rho_{\text{in}}} - f_2\frac{R_p^3}{2r^2}\nabla\cdot\left(\frac{\vec u_{\text{in}}}{r}\right)\mbox{ .}
\end{equation}
this potential actually satisfies a non-homogeneous  wave equation. By applying the D'Alembert operator, as it is done in \eqref{eq:F_sc}, the following source is gathered
\begin{equation}\label{eq:source_sc}
    \nabla^2\phi_{\text{sc}} - \frac{1}{c_0^2}\frac{\partial^2\phi_{\text{sc}}}{\partial t^2} = f_1\frac{V_p}{\rho_0}\dot{\rho_{\text{in}}}\delta(\vec r) + f_2\frac{3V_p}{2}\nabla\cdot\left(\vec u_{\text{in}}\delta(\vec r)\right)
\end{equation}
After plugging in \eqref{eq:source_sc} into \eqref{eq:F_sc},  we have
\begin{align}\label{eq:integrate_F_i}
    &\langle F_i \rangle = -V_p\int\left\langle f_1\dot{\rho_{\text{in}}}u_{\text{in}}^i\delta(\vec r) + f_2\frac{3\rho_0}{2}u_{\text{in}}^i\partial_k(u_{\text{in}}^k\delta(\vec r))\right\rangle dV \nonumber\\
    &= - f_2\frac{3\rho_0V_p}{2}\bigg( \oint \left\langle u_{\text{in}}^iu_{\text{in}}^k\delta(\vec r)\right\rangle dS_k \nonumber\\ 
    &- \int \left\langle (u_{\text{in}}^k\partial_k)u_{\text{in}}^i\delta(\vec r)\right\rangle dV \bigg) -f_1V_p\left\langle\frac{\partial\rho_{\text{in}}}{\partial t}u_{\text{in}}^i\right\rangle 
\end{align}
Because the Dirac's delta of the second term of does not contain the surface, the whole integrand is identically zero, leading to
\begin{equation}\label{eq:F_after_int}
    \langle F_i \rangle = -f_1V_p\left\langle\frac{\partial\rho_{\text{in}}}{\partial t}u_{\text{in}}^i\right\rangle + f_2\frac{3\rho_0V_p}{2}\left\langle(u_{\text{in}}^k\partial_k)u_{\text{in}}^i\right\rangle\mbox{ .}
\end{equation}
As a final step, we can exchange the time derivative in the first term, because the derivative of the whole product is identically zero (just because the incident field oscillates harmonically); thus
\begin{equation}
    \left\langle\frac{\partial\rho_{\text{in}}}{\partial t}u_{\text{in}}^i\right\rangle = \frac{1}{2\rho_0 c_o^2}\left\langle\partial_ip_{\text{in}}^2\right\rangle\quad,
\end{equation}
where \eqref{eq:linear_NSE_eqs} was considered. By using \eqref{eq:double_dev_property} in the second term of \eqref{eq:F_after_int} and by replacing the previous result, the Gor'kov Force takes its definitive form,
\begin{equation}\label{eq:F_grad_U_ap}
    \langle F_i \rangle = -\partial_i V_p\left(f_1\frac{1}{2\rho_0 c_0^2}\langle p_{\text{in}}^2 \rangle + f_2\frac{3\rho_0}{4} \langle u_{\text{in}}^2 \rangle\right) = -\nabla U
\end{equation}
written in \eqref{eq:F_grad_U}.

\section{Comments on the acoustic radiation force in viscous fluids}\label{ap:viscosity}

\begin{figure}
    \centering
    \includegraphics[width=0.5\textwidth]{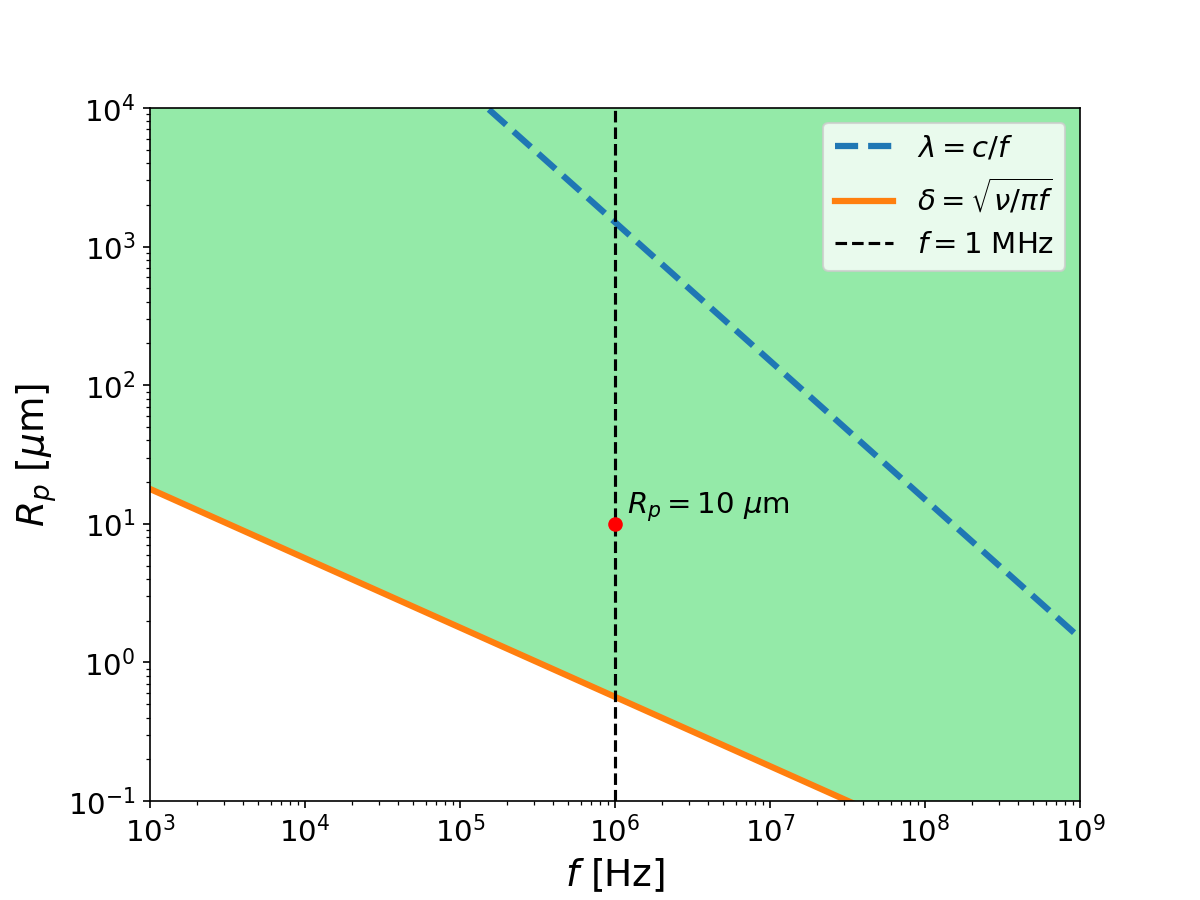}
    \caption{Viscosity can be neglected to compute the acoustic radiation force when the particle radius is larger than the penetration depth.}
    \label{fig:omega_lmd_delta}
\end{figure}

The case of a compressible sphere in a viscous fluid was first studied by Alexander A. Doinikov in 1994 \cite{m1_doinikov1994}.
He defined a viscous penetration depth as
\begin{equation}
    \delta = \sqrt{\frac{2\nu}{\omega}}    
\end{equation}
with $\nu$ the kinematic viscosity and $\omega$ the sound angular frequency.
For $\delta\ll R_p$ the viscosity effects can be neglected, thus, the dimensionless factor 
\begin{equation}
 \frac{\delta}{R_p} = \frac{\sqrt{\frac{2\nu}{cR_0}}}{\sqrt{kR_p}}   
\end{equation}
is useful to determine if viscosity must be included or not \cite[p.~3048]{n6_Marston2017}. In Fig.\ref{fig:omega_lmd_delta} we show the regimes for which the viscosity can be neglected (green shaded region). On the top of that we also report the constraint imposed by our linear expansion in $R_p/\lambda$ (dashed line). While our numerical approach is valid in the region between the lines, experimental results have shown that viscosity can be neglected in the full shaded area \cite{2n1_Herrey1955}. 
Additionally, S. D. Danilov and M. A. Mironov \cite{m2_danilov2000} showed that the particle can experience drag forces during its motion that may contribute significantly to the total motion dynamics of the sphere, but they do not affect the acoustic radiation force. Other viscous correction for the acoustic radiation force for spheres was proposed by M. Settnes and H. Bruus in 2012 \cite{m3_settnes2012} where the dipole coefficient $f_2$ is modified as follows:
\begin{equation}\label{eq:f_2_viscous}
    f_2 = \frac{2(\rho_p - \rho_0)(1 - \zeta(\delta))}{2\rho_p + \rho_0 - 3\rho_0\zeta(\delta)}\quad,
\end{equation}
where
\begin{equation}
    \zeta(\delta) = -\frac{3}{2}\left[1+i\left(1+\frac{\delta}{R}\right)\right]\frac{\delta}{R}\quad.
\end{equation}
The imaginary part of \eqref{eq:f_2_viscous} was corrected by L. P. Marston in 2016, showing in Eq. (13) of Ref. \cite{2n3_marston2016} that a factor of $2$ was missing in the Settnes and Bruus formula (Eq. (49) of Ref. \cite{m3_settnes2012}). For the case of cylinders, proposed corrections can be found in \cite{m4_Wang2011}, where acoustic forces and torques are calculated, and \cite{m5_Qiao2020} where the force on a cylinder close to a boundary in a viscous fluid is computed.

%

\end{document}